\documentclass[amssymb,twocolumn,aps,superscriptaddress]{revtex4-2}
\usepackage[utf8]{inputenc}
\usepackage{amsmath}
\usepackage{comment}
\usepackage{siunitx}
\usepackage{dcolumn}
\usepackage{braket}
\usepackage{chemformula}
\usepackage{graphicx,bm}
\usepackage{float}
\usepackage{subfigure}
\usepackage{amsmath}
\usepackage{amsfonts}
\usepackage{relsize}
\usepackage{tikz}
\usepackage{inputenc}
\usepackage{amsthm, amssymb,bbm,bm}

\usepackage{siunitx}
\usepackage{tabularx}
\usepackage{multirow}

\usepackage{physics}

\definecolor{myblue}{rgb}{0.18, 0.46, 0.67}
\usepackage{hyperref}
\hypersetup{colorlinks=true,breaklinks=true}
\AtBeginDocument{%
  \hypersetup{
    citecolor=myblue,
    linkcolor=myblue,   
    urlcolor=myblue}}

\begin{document}
\title{Long-range ballistic propagation of {80}$\%$-excitonic-fraction polaritons in a perovskite metasurface at room temperature}

\author{Nguyen Ha My Dang}\thanks{Nguyen Ha My Dang and Simone Zanotti contributed equally to this work as first authors} 
\affiliation{Universit\'e Lyon, Ecole Centrale de Lyon, CNRS, INSA Lyon, Universit\'e  Claude Bernard Lyon 1, CPE Lyon, CNRS, INL, UMR5270, Ecully 69130, France}

\author{Simone Zanotti\thanks{Nguyen Ha My Dang and Simone Zanotti contributed equally to this work as first authors. Corresponding authors:}}
\email{simone.zanotti01@universitadipavia.it} 
\affiliation{Dipartimento di Fisica, Universit\`{a} di Pavia, via Bassi 6, I-27100 Pavia, Italy}

\author{Emmanuel Drouard}
\affiliation{Universit\'e Lyon, Ecole Centrale de Lyon, CNRS, INSA Lyon, Universit\'e  Claude Bernard Lyon 1, CPE Lyon, CNRS, INL, UMR5270, Ecully 69130, France}

\author{C\'{e}line Chevalier}
\affiliation{Universit\'e Lyon, Ecole Centrale de Lyon, CNRS, INSA Lyon, Universit\'e  Claude Bernard Lyon 1, CPE Lyon, CNRS, INL, UMR5270, Ecully 69130, France}

\author{Ga\"{e}lle Tripp\'e-Allard}
\affiliation{Universit\'e Paris-Saclay, ENS Paris-Saclay, CentraleSup\'elec, CNRS, Lumi\`ere, Mati\`ere et Interfaces (LuMIn) Laboratory,  91190 Gif-sur-Yvette, France}

\author{Emmanuelle Deleporte}
\affiliation{Universit\'e Paris-Saclay, ENS Paris-Saclay, CentraleSup\'elec, CNRS, Lumi\`ere, Mati\`ere et Interfaces (LuMIn) Laboratory,  91190 Gif-sur-Yvette, France}

\author{Christian Seassal}
\affiliation{Universit\'e Lyon, Ecole Centrale de Lyon, CNRS, INSA Lyon, Universit\'e  Claude Bernard Lyon 1, CPE Lyon, CNRS, INL, UMR5270, Ecully 69130, France}

\author{Dario Gerace} 
\affiliation{Dipartimento di Fisica, Universit\`{a} di Pavia, via Bassi 6, I-27100 Pavia, Italy}

\author{Hai Son Nguyen}\email{hai-son.nguyen@ec-lyon.fr} 
\affiliation{Universit\'e Lyon, Ecole Centrale de Lyon, CNRS, INSA Lyon, Universit\'e  Claude Bernard Lyon 1, CPE Lyon, CNRS, INL, UMR5270, Ecully 69130, France}
\affiliation{Institut Universitaire de France (IUF), 75231 Paris, France} 
\begin{abstract}{
Exciton-polaritons, hybrid light-matter excitations arising from the strong coupling between excitons in semiconductors and photons in photonic nanostructures, are crucial for exploring the physics of quantum fluids of light and developing all-optical devices. Achieving room temperature propagation of polaritons with a large excitonic fraction is challenging but vital, e.g., for nonlinear light transport. We report on room temperature propagation of exciton-polaritons in a metasurface made from a sub-wavelength lattice of perovskite pillars. The large Rabi splitting, much greater than the optical phonon energy, decouples the lower polariton band from the phonon bath of perovskite. These cooled polaritons, in combination with the high group velocity achieved through the metasurface design, enable long-range propagation, exceeding hundreds of micrometers, even with an {80}$\%$ excitonic component. Furthermore, the design of the metasurface introduces an original mechanism for uni-directional propagation through polarization control, suggesting a new avenue for the development of advanced polaritonic devices.}
\end{abstract}

\maketitle
\section{Introduction}

{Exciton-polaritons are hybrid light-matter excitations emerging from the strong coupling regime between photons and excitons in semiconductors~\cite{art:Hopfield,Agranovich1960}.
Manipulating these excitations in confining geometries, like quantum wells embedded in microcavities~\cite{art:weisbuch}, has paved the way for probing out-of-equilibrium Bose-Einstein condensation and the rich physics of quantum fluids of light~\cite{QuantumFluidOfLight}.}
Exciton-polaritons are currently at the forefront of developing advanced all-optical devices~\cite{RoadmapPol}. Investigating polariton propagation is crucial for designing devices that fully exploit high-speed and efficient polaritonic signal transmission. Thanks to their hybrid nature, polariton  propagation features unique properties not found in purely photonic or excitonic transport. From their photonic component, exciton-polaritons benefit from a small effective mass and high group velocity, enabling ballistic propagation over macroscopic distances, and allowing for tailoring transport properties through the engineering of potential landscape~\cite{art:pol_trans,Ballarini2013a}. In addition, their excitonic component introduces highly nonlinear behaviors~\cite{art:non_linear_scatt}, giving rise to solitons~\cite{art:amo_sol}, nonlinear tunneling effects~\cite{art:hai_tun}, and even superfluidity~\cite{art:superfl,art:hai_BH}. The polariton flow can also be directed by external fields interacting with their excitonic component~\cite{art:full_guided}. 

A wide range of excitonic materials  has been utilized to study polariton propagation, including GaAs-based quantum wells for foundational research at cryogenic temperatures~\cite{art:wertz,art:steger,art:freixanet,art:langbein,art:surez,art:Walker,Tanese2012,art:Marsault,Ballarini2013a}, and materials that operate at room temperature such as ZnO~\cite{art:Franke,art:Ghosh_ucavityRT}, organic materials~\cite{art:high_speed,arT:Hou,art:Zhang}, GaN~\cite{art:Ciers}, transition-metal dichalcogenides (TMDs)~\cite{art:barachati,Liu2023,art:Wurdak_prop}, and perovskites~\cite{art:su,Peng2022,Xu2023,Chen2023,art:peng_RoomT_per,art:Soci_pero}. Nevertheless, much of this research has focused on photonic-like polaritons with a low excitonic fraction (less than 50$\%$). Overcoming the challenges associated with achieving macroscopic propagation of polaritons with a high excitonic fraction (i.e., larger than 50$\%$) is essential to harness the full potential of these hybrid excitations in communication devices. {Long-range polariton propagation with a high excitonic fraction is crucial for exploiting polariton nonlinearity in all-optical devices operating at low threshold when using polaritons as information carriers. Moreover, nonlinear transport of polaritons can unlock new paradigms for information transmission in next generation devices, such as Anderson-localization-free~\cite{Tanese2012} or superfluid~\cite{art:superfl,art:hai_BH} propagation.} The primary obstacle is that the microcavity design,commonly used for generating exciton-polaritons, offers relatively low group velocities, which decrease significantly as the excitonic fraction increases. Additionally, the thermal broadening of the excitonic resonance at room temperature poses a significant challenge for the  propagation of highly excitonic polaritons. For example, one of the most used perovskite materials, (C$_6$H$_5$C$_2$H$_4$NH$_3$)$_2$PbI$_4$ (PEPI), displays a strong excitonic resonance with a \SI{30}{\milli\electronvolt} {homegeneous} linewidth 
at room temperature~\cite{art:neutzner,Kandada2022}. This suggests that PEPI-based polaritons with an excitonic fraction above 50$\%$ would have a minimum linewidth of \SI{15}{\milli\electronvolt}, {even in the best case scenario of negligible photonic losses}. For a group velocity of \SI[per-mode = symbol]{2}{\micro\meter\per\pico\second}, which is typical of polaritons in microcavity samples, this linewidth corresponds to a propagation length of less than \SI{100}{\nano\meter}.

{In this work, we overcome these limitations by}  engineering a photonic potential landscape, tailored through a sub-wavelength lattice of pillars resonant with the excitonic resonance of the PEPI perovskite.  {This dispersion engineering allows for the decoupling of high-speed and excitonic-like polaritons from the thermal bath, ultimately enabling polariton propagation over macroscopic distances at room temperature. As a result, ballistic propagation of polaritons with {80}$\%$ excitonic fraction at a speed of \SI[per-mode = symbol]{25}{\micro\meter\per\pico\second} and a linewidth as small as a few meV is experimentally demonstrated across distances exceeding 100 µm. Our findings suggest an original mechanism for long distance polariton propagation, and provide a unique platform to engineer polaritonic transport at room temperature.}

\section{Polariton eigenmodes from a large nano-imprinted perovskite metasurface}

\begin{figure}
\centering
    \includegraphics[width=1\linewidth]{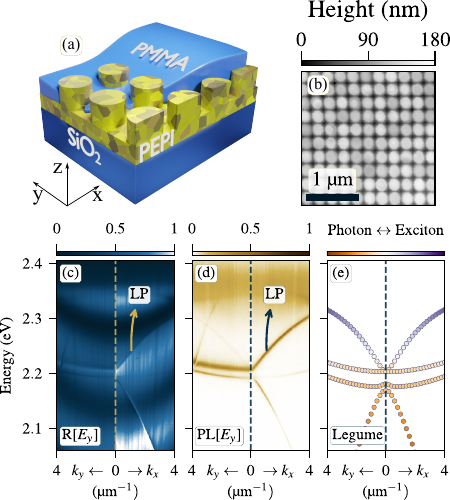}
    \caption{(a) Three-dimensional sketch of the active metasurface studied in this work ({not to scale}). (b) Atomic force microscope image of the sample. (c) Reflectivity ($R$) and (d) photoluminescence ($PL$) spectra for $E_y$-polarized light with wavevector pointing along the $k_x$ and $k_y$ directions, respectively, and obtained from the structure pictured in (a). In this paper, our focus lies on the study of the propagation from the lower polariton branch, referred to as ``LP'' in panels (c)-(d). The broadband signal that is visible in (d) around \SI{2.394}{\electronvolt} is attributed to bare PEPI exciton emission. {(e) First principles calculated polariton dispersion, the colorscale encodes the excitonic fraction.}}  
    \label{fig:struct}
\end{figure}

{Our sample consists of a} high-quality and homogeneous PEPI metasurface, covering an  area {of $\sim$ 3 cm$^2$}. This metasurface was fabricated using the thermal imprinting method~\cite{Mermet2023,art:dang2024} {(fabrication details are reported in the Supporting information (SI) and detailed in  Ref.~~\cite{art:dang2024})}. 
{
A sketch of the fabricated metasurface is shown in Fig.~\ref{fig:struct}(a). The overall PEPI thickness is estimated as \SI{93}{\nano\meter} with a  patterned thickness of $t_{\text{pattern}} = \SI{35}{\nano\meter}$ and an unpatterned thickness $t_{\text{slab}} = \SI{58}{\nano\meter} $. The metasurface consists of a square lattice of pillars, with a period of $a=\SI{295}{\nano\meter}$ and a diameter of \SI{250}{\nano\meter} (see  Fig.~\ref{fig:struct}(b)). }

Signatures of polariton excitations in the system are experimentally probed by angle-resolved reflectance (ARR) and photoluminescence (ARPL), respectively (see SI for experimental setup details). Signals for both measurements can be analyzed in two polarizations: $E_y$ (S-polarization with respect to $xz$ plane) and $E_x$ (P-polarization with respect to $xz$ plane).  Fig.~\ref{fig:struct}(c,d) shows results for  $E_y$ polarization along both $k_x$ and $k_y$. Due to the C$_4$ symmetry of the square lattice, $E_x$ polarization results along $k_x$ and $k_y$  are identical to those for $E_y$ polarization along $k_y$ and $k_x$, respectively.
{ Fig.~\ref{fig:struct}(e) compares the experimental spectra with the polariton bands calculated using \texttt{legume}, an open-source software for first-principles calculations of polariton dispersion and excitonic fraction~\cite{art:Zanotti2024,repo:legume}. The simulation including both $E_x$ and $E_y$ polarized modes, shows a good match for each polariton mode. More simulation details are given in the SI.}

{We focus on the LP (Lower Polariton) mode in Figs.~\ref{fig:struct}(c,d). The strong coupling regime is clearly evidenced by the anticrossing effectin ARR, ARPL spectra {and calculated dispersion},  highlighted by the bending of the LP mode as it approaches the exciton energy  $E_X=\SI{2.394}{\electronvolt}$. {This bending is even more pronounced in ARPL measured from the tilted sample, showing emission at wave-vectors up to 7 µm$^{-1}$ (see SI)}.  The LP mode's dispersion is almost flat along $k_y$ but highly dispersive along $k_x$, with a relatively large group velocity, making it ideal for studying polariton propagation along the $x$ direction. {In addition, the C$_4$ symmetry imposes a similar polaritonic mode of $E_x$ polarization for studying propagation along the $y$ direction.  }

\begin{figure}
\centering
    \includegraphics[width=1\linewidth]{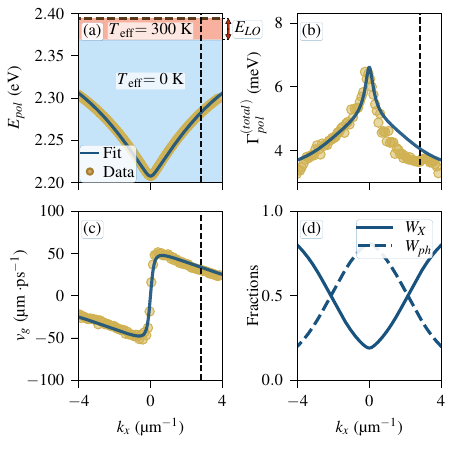}
    \caption{The {yellow dots in panels (a), (b) and (c) represent the} energy peak position, the linewidth, and the group velocity,  respectively, measured for the polaritonic branch (LP in Fig.~\ref{fig:struct}), as a function of the wave vector, $k_x$. The full lines in (a), (b), and (c) are theoretical curves for the LP polaritonic mode obtained from the coupled oscillators model, with the excitonic resonance indicated as a dashed line in (a). From the same model, we plot in (d) the Hopfield coefficients, the dashed (full) line representing the calculated photonic (excitonic) fraction of this polaritonic mode as a function of the wave vector. In (a,b,c), the vertical dashed lines correspond to the wave vector for which polaritons possess excitonic fraction {$W_X=63\%$, corresponding to the propagation measurements in Fig.~\ref{fig:3} and Fig.\ref{fig:decay} .}  }
    \label{fig:fit}
\end{figure}

Figures~\ref{fig:fit}(a,b,c) present the resonance energy, the full-width at half-maximum (FWHM) linewidth, and group velocity of LP, extracted from ARPL {measurements}. The {experimental} results are nicely fitted by the coupled oscillator model {between PEPI excitons and photonic Bloch resonances, given by}:
\begin{equation}\label{eq:Coupled_oscillator}
    \tilde{E}_{pol}(k_x) = \frac{\tilde{E}_{ph}(k_x)+\tilde{E}_X}{2}-\frac{1}{2}\sqrt{4V^2+ \left[\tilde{E}_{ph}(k_x)-\tilde{E}_X\right]^2  } \, .
\end{equation} 
Here, we can explicitly separate the real part contribution (i.e., the resonances dispersion) from the losses due to the imaginary part: {$\tilde{E}_{pol}(k_x) =E_{pol}(k_x)+i\Gamma_{pol}(k_x)/2 $.
In the expression given in Eq.~\eqref{eq:Coupled_oscillator}, $V$ is the exciton-photon coupling strength, $\tilde{E}_X=E_X + i\Gamma_X/2$ is the dispersionless excitonic band of energy $E_X$ and linewidth $\Gamma_X$, and $\tilde{E}_{ph}(k_x)$ is the complex photonic mode dispersion, which can be approximated by an analytic expression as~Eq.\eqref{eq:phot_disp} {(see SI for expression of all photonic modes along both $k_x$ and $k_y$)}:
\begin{equation}\label{eq:phot_disp}
\tilde{E}_{ph}(k_x) = E_0+i\frac{\Gamma_0}{2}+ \sqrt{\left(\hbar vk_x\right)^2+\left(U+i\frac{\Gamma_0}{2}\right)^2 } \, .
\end{equation} }
In Eq.~\eqref{eq:phot_disp}, $v$ is the group velocity of the guided modes, which are then folded and coupled by the periodic metasurface. The coupling strength between guided modes is given by the parameter $U$, while $\Gamma_0$ accounts for the coupling between guided modes and the radiative continuum, due to the periodic dielectric modulation. From Eq.~\eqref{eq:phot_disp}, one may extract the photonic dispersion, $E_{ph}(k_x)$, as well as the photonic losses, $\Gamma_{ph}(k_x)$, by calculating the real and imaginary parts of $\tilde{E}_{ph}(k_x)$, respectively. 

{The group velocity along $x$  is  obtained from the derivative of the real part of the polaritonic dispersion.  In the wavevector range shown in Fig.~\ref{fig:fit}(c), the group velocity rapidly increases from $v_g=\SI{0}{\micro\meter\per\pico\second}$ to $v_g=\SI{50}{\micro\meter\per\pico\second}$ with a slight wavevector increase; then remains above $v_g=\SI{25}{\micro\meter\per\pico\second}$ at larger wavevectors. }
 
 Figure~\ref{fig:fit}(d) shows the calculated Hopfield coefficients (i.e., excitonic and photonic fractions, respectively), given by:
 \begin{equation}\label{eq:Hopfield_coeff}
 W_X, W_{ph} \approx \frac{1}{2}\left(1\pm \frac{E_{ph}-E_X}{\sqrt{(E_{ph}-E_X)^2 + 4V^2}}\right).
 \end{equation}
 {Our results demonstrate that perovskite metasurfaces can exhibit highly excitonic polaritons with high group velocity (tens of \si{\micro\meter\per\pico\second}) and low losses (few \si{\milli\electronvolt}). }
 
 {For model parameters, the exciton energy is extracted from the absorption measurement on an unpatterned PEPI thin film, resulting in $E_X=\SI{2.394}{\electronvolt}$ {(see SI)}.  From fitting, we estimated { $E_0=\SI{2.294}{\electronvolt}$, $v=\SI{77.5}{\micro\meter\per\pico\second}$, $U=\SI{10} {\milli\electronvolt}$, $\Gamma_0=\SI{2.2} {\milli\electronvolt} $ and $\Gamma_X= \SI{0.3} {\milli\electronvolt}$. The exciton-photon coupling strength is $V=\SI{127}{\milli\electronvolt}$, leading to a Rabi-splitting energy of $\sim \SI{254}{\milli\electronvolt}$ at zero detuning, agreeing with previous reports~\cite{Dang2020,art:HaMy2022AOM}. An inhomogeneous broadening of $\Gamma_{inh}=\SI{3}{\milli\electronvolt}$ is added to the total polariton linewidth: $\Gamma_{pol}^\text{(total)}=\Gamma_{pol} + \Gamma_{inh}$}. These parameters accurately reproduce the complex LP mode dispersion, as shown in Figs.~\ref{fig:fit}(a,b,c), and in iso-frequency measurements and propagation experiments in both real and momentum space, which will be presented in the following. 

\section{Suppression of thermal broadening of perovskite polaritons }\label{sec:suppress}

Intriguingly, the excitonic losses estimated from the fit in the previous Section,{ $\Gamma_X=0.3$ meV}, are orders of magnitude smaller than the value ($\sim$\SI{30}{\milli\electronvolt}) that is either {reported} in the literature for {the homogeneous linewidth of PEPI excitons at room temperature~\cite{art:neutzner,Kandada2022}}, or obtained from a rough estimate of the FWHM {of uncoupled excitons in the PL spectra of Fig.~\ref{fig:struct}(d)}. {Moreover, as shown in Fig.~\ref{fig:fit}(b), the polariton linewidth evidently narrows down as the excitonic fraction increases.} As a direct consequence, at $63\%$ excitonic fraction (i.e., $k_x\approx\SI{2.8}{\per\micro\meter}$), the measured polariton linewidth is about $\SI{4}{\milli\electronvolt}$. {Additionally,} from the propagation experiment described in the next section, we extract a polariton lifetime of 0.7 ps,  corresponding to a homogeneous linewidth of 0.94 meV for polaritonic states with $63\%$ excitonic fraction. 
This unusual narrowing of the polariton linewidth is explained by the suppression of the thermal broadening for our polaritonic states, as thoroughly described in the following.

{In general}, exciton-photon interaction is orders of magnitude stronger than exciton-phonon coupling~\cite{art:Savona-Piermarocchi1997}. Thus, polaritons are primarily formed from photons and  ``isolated'' excitons. {Subsequently}, the thermal broadening is due to polariton-phonon interaction and not to exciton-phonon interaction. As a consequence, the polariton {homogeneous} linewidth is {given by}: 
\begin{equation}
    \Gamma_{pol}\left( T\right)=  \underbrace{W_{ph}\cdot\Gamma_{ph}+W_X\cdot\Gamma_X}_\text{coupled-oscillator linewidth}+\Gamma_{\text{th} }\left(T\right) \, ,
\end{equation}
{in which $\Gamma_X$ is the {homogeneous exciton linewidth} at $\SI{0}{\kelvin}$ (notice, not at the temperature $T$!),} and $\Gamma_{\text{th} }\left(T\right)$ is the thermal broadening induced from the polariton-phonon interaction. 

The main contribution to polaritonic dephasing at high-temperature is the optical phonon-polariton interaction, which scatters polaritons into excitonic states~\cite{art:trichet,trichetPhD,Ferreira2022}. This process is favored by the high density of states in the excitonic reservoir at $E_X$. Energy conservation allows such scattering in the range $E_{pol}\in \left[E_X-E_{\text{LO}},E_X\right]$, where $E_{\text{LO}}\sim 20-30$\,meV is the optical phonons energy in PEPI~\cite{art:gauthron,art:neutzner,Kandada2022,Feldstein2020}. This  divides the spectrum into two regions, {as represented  in Fig.~\ref{fig:fit}(a)}: 1) $E_{pol} > E_X-E_{\text{LO}}$, where polaritons scatter efficiently with phonons at room temperature; 2) $E_{pol}<E_X-E_{\text{LO}}$, where the phonon-polariton scattering is suppressed, and polaritons effectively behave as if in an environment at $T_{\text{eff}}=\SI{0}{\kelvin} $. {Strikingly, all the experimental data for the LP band align with these polaritons being in the $T_{\text{eff}}=\SI{0}{\kelvin}$ region, regardless of excitonic fraction. Therefore, these polaritons are ``cooled" and do not exhibit any thermal broadening: $\Gamma_{th}\approx0$}. This argument gives a remarkably simple but effective account for the reduced polariton linewidth observed in the experimental data in {Fig.~\ref{fig:fit}(b)}.

{We further notice that the suppression of the polariton-phonon scattering channel was theoretically suggested and experimentally demonstrated by Trichet et al.~\cite{art:trichet,trichetPhD} for  ZnO-based polaritons. A similar mechanism was recently predicted for TMD-based polaritons.~\cite{Ferreira2022}.} {In the hypothesis of low losses ($\Gamma_{pol} \ll E_{pol} $, $ \Gamma_{X} \ll E_{X}$), the condition $E_{pol}<E_X - E_{\text{LO}} $ can be rewritten from Eq.\eqref{eq:Coupled_oscillator} and Eq.\eqref{eq:Hopfield_coeff} as:}
\begin{equation}\label{eq:cooled_pols}
    V>E_{\text{LO}}\sqrt{\frac{W_X}{1-W_X}} \, .
\end{equation}
 For $W_X=50\%$, this condition corresponds to $V>E_{\text{LO}}$~\cite{art:trichet,trichetPhD} which is difficult to be satisfied for exciton-polaritons in III-V semiconductors or TMD monolayers, but {easily} met in materials with stronger oscillator strengths, such as PEPI. Thus, for PEPI polaritons we can safely assume that $\Gamma_{th}\approx0$, even for polariton eigenmodes with large excitonic fraction. {Specifically, for our PEPI metasurface with $V=\SI{127}{\milli\electronvolt}$ and $E_{\text{LO}}=\SI{30}{\milli\electronvolt}$, the cooled polaritons condition is satisfied for $W_X<95\%$.} 
 {We stress that the condition given by Eq.~\eqref{eq:cooled_pols} is general, and applies to any cavity geometry and excitonic material. Reports on perovskite polaritons in planar microcavities have shown hints of similar linewidth narrowing in PL spectra but have overlooked this effect~\cite{Bao2019, Wang2018}. However, careful analysis of polaritonic mode narrowing is needed, as it may result from various mechanisms.  Thermal broadening suppression is dominant if the bare-exciton linewidth is mainly homogeneous, as in ZnO micro-wires or perovskites at room temperature. If the bare-exciton linewidth is dominated by inhomogeneous broadening, as in epitaxially grown III-V quantum wells at cryogenic temperatures or TMD monolayers and molecular vibrations at room temperature, the narrowing is governed by the motional narrowing mechanism~\cite{Whittaker1996, Savona1997, Kravtsov2020, Wurdack2021, Verdelli2024}.}

\section{Polariton propagation: measurements in real space}
In this section, we present the results of a propagation experiment involving polaritonic eigenmodes with $\sim63\%$ excitonic fraction. From the PL signal under non-resonant pumping, this fraction is selected by using a spectral band-pass filter centered at $E_{f}=\SI{2.283}{\electronvolt}$ {(see SI)}. This spectral filtering corresponds to an average wave vector of \SI{2.8}{\per\micro\meter}, with an excitonic fraction of $63\%$.
\begin{figure}
\centering
    \includegraphics[width=1\linewidth]{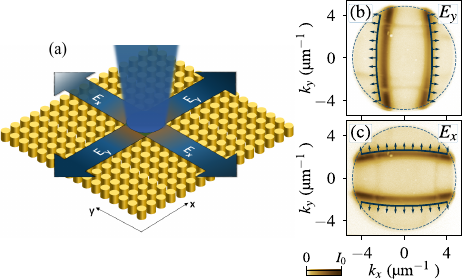}
    \caption{(a) Schematic representation of the in-plane propagation of polaritonic modes excited with $E_x$ and $E_y$ polarised light, respectively. (b)-(c) Iso-frequency wavevector dispersion of exciton-polaritons, {filtered at $E_0=\SI{2.283}{\electronvolt}$}, as obtained from mapping the far-field emission in (b) $E_y$ and (c) $E_x$  polarisations, respectively. The arrows indicate the group velocity patterns. The dashed circles indicate the numerical aperture of the objective lens, while the solid lines are the iso-frequency cuts of the bands calculated with Eq.~\eqref{eq:Coupled_oscillator}, using the same parameters as in Fig.~\ref{fig:fit}. {Details on the {theoretical} model for the description of polaritonic modes, both as a function of $k_x$ and $k_y$, are given in the {SI.}}}
    \label{fig:3}
\end{figure}
Since the propagation direction is dictated by the group velocity vectors, it is crucial to investigate the group velocity pattern  at the selected energy before studying the propagation itself. This pattern can be visualized from the iso-frequency mapping of the polaritonic modes. Indeed, the group velocity follows the normal direction of the iso-frequency curves, as defined by the gradient of the energy surface in momentum space: {$\mathbf{v}_g = 1/\hbar\, \nabla_{  \mathbf{k}}E_{pol}$}. Figure~\ref{fig:3}(b)-(c) presents the measured iso-frequency map for two different polarizations, obtained from the far-field PL emission. Two polaritonic modes are revealed: one $E_y$-polarized, with dispersion along $k_x$ (the {LP} mode shown in Fig.~\ref{fig:struct}(c)-(d)), and the other $E_x$-polarized, with dispersion along $k_y$. {The theoretical calculations are also plotted, showing very good agreement with experimental results.}
The iso-frequency cuts show that the $E_y$-($E_x$-)polarized mode contains two approximately parallel {bands} along $k_y$ ($k_x$), with a group velocity pattern that is almost uniform along the $x$ ($y$)  direction. {This is confirmed by the measurement of the group velocity of the LP mode for different values of $k_y$, as shown in the SI}.

From the results above, we expect polariton propagation in the 2D metasurface to consist of two one-dimensional flows of uniform propagating fronts with constant group velocity along the $x$ and $y$ directions, with orthogonal polarizations, as  depicted in Fig.~\ref{fig:3}(a). This is experimentally demonstrated by visualizing the farfield image in real space. Fig.~\ref{fig:decay}(a) presents the spectrally filtered PL image in real space, measured in $E_y$ polarization. In this experiment, the non-resonant pumping of $\sim 2-3$ \si{\micro\meter} diameter is focused at $x=y=0$. These results clearly show that polaritons are locally injected at $x=y=0$, and  propagate along the $x$ direction. Strikingly, the polariton flow remains tightly focused along $y$ direction, even about $x=\SI{60}{\micro\meter}$ away from the pumping spot, as evident in Fig.~\ref{fig:decay}(a). This observation confirms that the velocity vector is aligned along the $x$ direction. Evidently, by switching polarization from $E_y$ to $E_x$, also the propagation direction switches from $x$ to $y$.

\begin{figure}
\centering
    \includegraphics[width=1\linewidth]{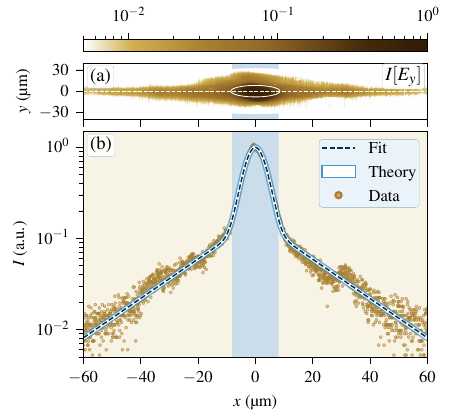}
    \caption{(a) Spatially resolved photoluminescence image of the polaritonic states filtered at \SI{2.283}{\electronvolt}, with $E_y$ polarization selection. The blue-shaded area in the centre corresponds to the pump spot region. (b) Intensity profile selected at $y=0$. The scattered points represent the experimental data, {while the dashed line is the fitting results obtained from Eq.~\eqref{eq:decay}, and the white line with blue contour is calculated with the theoretical model for the polariton propagation presented in the SI}.}
    \label{fig:decay}
\end{figure}

To quantitatively evaluate the propagation properties, the intensity profile at $y=0$ is extracted from the PL image and reported in Fig.~\ref{fig:decay}(b). In the real-space PL, there is some PEPI uncoupled excitonic signal around the excitation spot. Still, these excitons possess high effective mass and zero group velocity, preventing long-distance propagation. Thus, the PL signal detected far away from the pumping spot comes from propagating polaritons. The PL intensity is fitted using a phenomenological function: a Gaussian for the excitonic reservoir and an exponential decay for polariton propagation:
\begin{equation}\label{eq:decay}
    I(x)=I_0e^{-\frac{x^2}{2\sigma^2}}+I_1e^{-\frac{|x|}{l}} \, .
\end{equation}
Here, $\sigma$ is the size of the pumping spot and $l$ is the decay length. From the fit, we estimated a propagation constant $l= \SI{20}{\micro\meter}$ and a pumping spot size $\sigma\sim\SI{3.5}{\micro\meter} $. This value of decay length corresponds to a polariton lifetime of \SI{0.7}{\pico\second}, assuming ballistic propagation. This lifetime corresponds to a homogeneous linewidth of \SI{0.94}{\milli\electronvolt}. We notice that this linewidth is much smaller than the \SI{4}{\milli\electronvolt} value obtained from the previous section's fit. This indicates that the inhomogeneous broadening effects are the dominant contributions to the total polariton linewidth in the emission spectra. {This observation aligns with the $\Gamma_{inh}=3$ meV used as a fitting parameter for inhomogeneous broadening in the results of Fig.~\ref{fig:fit}. Finally, the same parameters for the  polaritonic modes are implemented in a 1D propagation model (see SI). The calculated spatial decay from this model, shown in Fig.~\ref{fig:decay}, perfectly reproduces the experimental results and the phenomenological law.} 

\section{Polariton propagation: measurements in momentum space}
\begin{figure}
\centering
    \includegraphics[width=1\linewidth]{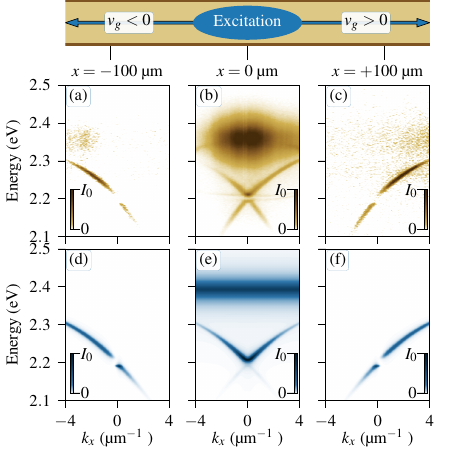}
    \caption{(a)-(b)-(c) Normalized PL spectra obtained by spatially filtering the signal at $x=\SI{-100}{\micro\meter}$, $x=\SI{0}{\micro\meter}$ and $x=\SI{+100}{\micro\meter}$, respectively. The pump spot is focused at $x=\SI{0}{\micro\meter}$. (d)-(e)-(f) Simulated PL spectra calculated for the same positions as (a)-(b)-(c). Details on the {theoretical} model for the description of polariton propagation, {as well as on the signal denoising,} are given in the {Supporting Material}. 
    }
    \label{fig:propagation}
\end{figure}

To investigate the ballistic nature of  polariton propagation, we monitor this propagation in momentum space to probe possible backscattering. A slit of width $S=\SI{17}{\micro\meter}$, acting as a spatial filter, is placed at the intermediate image position in the optical path of the ARPL setup~\cite{Cueff2024}. By moving the slit along the $x-$axis, we can probe the polariton dispersion either immediately below or far from the excitation spot. In this experiment, the spectral filter from the previous experiment is removed. The experimental results for three different slit positions are shown in Fig.~\ref{fig:propagation}(a-c), with corresponding theoretical calculations from the 1D propagation model shown in Fig~\ref{fig:propagation}(d-e). }

Under the pumping spot at $x=0$, the ARPL results show all of the polaritonic branches that were previously visible in the reflectivity and PL spectra of Fig.~\ref{fig:struct}(c)-(d). The branches in Fig.~\ref{fig:propagation} are broadened compared to those measured without a spatial filter in Fig.~\ref{fig:struct}(c)-(d) due to an uncertainty on the emission wavevector, {$\Delta_k\approx 1/S$}, introduced by the spatial filter's finite extent. Moreover, the intensity distribution is very different that measured without the spatial filter: the strongest signals are observed from the uncoupled excitons and the zero-group velocity point in the polariton branches. Indeed, polaritons/excitons injected into these states cannot propagate, and radiatively emit photoluminescence under the pump spot, while states with non-zero group velocity propagate out of the pump spot region.
 
{The ballistic nature of polaritonic propagation in our system is fully supported by the absence of back-scattering signals in momentum-space measurements, as seen in Figs.~\ref{fig:propagation}(a) and \ref{fig:propagation}(c). Only polaritons with positive group velocity are observed at $x>0$, and  polaritons with negative group velocity  at $x<0$. This is in contrast with the long-range propagation of polaritons in disordered organic semiconductors, where transport is dominated by scattering, resulting in strong back-scattering signals~\cite{Hou2020}}.

Finally, the absence of the signal from uncoupled excitons far from the pump spot allows us to observe the linewidth narrowing of the LP branch as it approaches the excitonic energy. This further confirms our interpretation of the quenching of the thermal broadening. Impressively, this propagation across \SI{100}{\micro\meter} is observed even with polaritons at $k_x=\SI{4}{\per\micro\meter}$, corresponding to an excitonic fraction of {80}$\%$. Since the extracted lifetime for polaritonic eigenmodes with a $63\%$ excitonic fraction is \SI{0.7}{\pico\second}, limited only by the photonic losses, we estimate that the lifetime of these {80}$\%$ excitonic fraction polaritons is even longer, about \SI{1.3}{\pico\second}. 
{We highlight that our polaritonic propagation is  unrelated from exciton diffusion in bare-perovskite layers. The latter is quantified in a few hundred nanometers and strongly dictated by exciton-phonon interactions~\cite{Ziegler2020, Seitz2020, Xiao2020}. Discussions on the diffusion of uncoupled excitons in our system are reported in the SI.}

\section{Conclusion and Perspectives}

We have demonstrated  ballistic propagation of high-excitonic-fraction polaritons over the hundred-micrometer-range at room temperature. This achievement is enabled by suppressing thermal-induced broadening in perovskite polaritons, formed within a large-scale and homogeneous perovskite metasurface fabricated through direct nano-imprint technology. The metasurface design introduces an innovative mechanism for directing unidirectional and high-speed propagation across macroscopic distances through polarization control. Given the strong nonlinearity exhibited by PEPI polaritons at high excitonic fractions~\cite{Fieramosca2019}, our platform opens new avenues for exploring nonlinear transport of quantum fluids of light beyond traditional microcavity architectures. By leveraging precise tailoring of photonic dispersion~\cite{Nguyen2018} or employing novel concepts of loss engineering like bound states in the continuum~\cite{art:zanotti,Nigro2023,Sigurdsson2024}, we anticipate unlocking new regimes of superfluidity and non-equilibrium hydrodynamics~\cite{art:peng_RoomT_per}. Finally, the ability to route and confine low-loss, highly nonlinear polaritons at room temperature using a scalable, cost-effective fabrication method paves the way for integrating polariton physics into all-optical or even electrically-driven~\cite{art:Soci_pero,art:Elettr_Pump,art:Wang_El_Pump} and integrated devices.
\vspace{2cm}

{\textit{Acknowledgement:} The authors would like to thank the staff from the Nanolyon Technical Platform for helping and supporting all nanofabrication processes. We also thank Le Si Dang and Lucio Claudio Andreani for fruitful discussions. H.S.N, C.S, E.D acknowledge financial support by the French National Research Agency (ANR) under the project POPEYE (ANR-17-CE24-0020), project EMIPERO (ANR-18-CE24-0016). S.Z. and D.G. acknowledge financial support from PNRR MUR project PE0000023-NQSTI. }


\bibliography{main}

\begin{thebibliography}{69}%
\makeatletter
\providecommand \@ifxundefined [1]{%
 \@ifx{#1\undefined}
}%
\providecommand \@ifnum [1]{%
 \ifnum #1\expandafter \@firstoftwo
 \else \expandafter \@secondoftwo
 \fi
}%
\providecommand \@ifx [1]{%
 \ifx #1\expandafter \@firstoftwo
 \else \expandafter \@secondoftwo
 \fi
}%
\providecommand \natexlab [1]{#1}%
\providecommand \enquote  [1]{``#1''}%
\providecommand \bibnamefont  [1]{#1}%
\providecommand \bibfnamefont [1]{#1}%
\providecommand \citenamefont [1]{#1}%
\providecommand \href@noop [0]{\@secondoftwo}%
\providecommand \href [0]{\begingroup \@sanitize@url \@href}%
\providecommand \@href[1]{\@@startlink{#1}\@@href}%
\providecommand \@@href[1]{\endgroup#1\@@endlink}%
\providecommand \@sanitize@url [0]{\catcode `\\12\catcode `\$12\catcode `\&12\catcode `\#12\catcode `\^12\catcode `\_12\catcode `\%12\relax}%
\providecommand \@@startlink[1]{}%
\providecommand \@@endlink[0]{}%
\providecommand \url  [0]{\begingroup\@sanitize@url \@url }%
\providecommand \@url [1]{\endgroup\@href {#1}{\urlprefix }}%
\providecommand \urlprefix  [0]{URL }%
\providecommand \Eprint [0]{\href }%
\providecommand \doibase [0]{https://doi.org/}%
\providecommand \selectlanguage [0]{\@gobble}%
\providecommand \bibinfo  [0]{\@secondoftwo}%
\providecommand \bibfield  [0]{\@secondoftwo}%
\providecommand \translation [1]{[#1]}%
\providecommand \BibitemOpen [0]{}%
\providecommand \bibitemStop [0]{}%
\providecommand \bibitemNoStop [0]{.\EOS\space}%
\providecommand \EOS [0]{\spacefactor3000\relax}%
\providecommand \BibitemShut  [1]{\csname bibitem#1\endcsname}%
\let\auto@bib@innerbib\@empty
\bibitem [{\citenamefont {Hopfield}(1958)}]{art:Hopfield}%
  \BibitemOpen
  \bibfield  {author} {\bibinfo {author} {\bibfnamefont {J.~J.}\ \bibnamefont {Hopfield}},\ }\bibfield  {title} {\bibinfo {title} {Theory of the contribution of excitons to the complex dielectric constant of crystals},\ }\href {https://doi.org/10.1103/PhysRev.112.1555} {\bibfield  {journal} {\bibinfo  {journal} {Phys. Rev.}\ }\textbf {\bibinfo {volume} {112}},\ \bibinfo {pages} {1555} (\bibinfo {year} {1958})}\BibitemShut {NoStop}%
\bibitem [{\citenamefont {Agranovich}(1960)}]{Agranovich1960}%
  \BibitemOpen
  \bibfield  {author} {\bibinfo {author} {\bibfnamefont {V.~M.}\ \bibnamefont {Agranovich}},\ }\bibfield  {title} {\bibinfo {title} {Dispersion of electromagnetic waves in crystals},\ }\href {https://doi.org/10.1038/s41377-020-0286-z} {\bibfield  {journal} {\bibinfo  {journal} {Sov. Phys. JETP}\ }\textbf {\bibinfo {volume} {37}},\ \bibinfo {pages} {307} (\bibinfo {year} {1960})}\BibitemShut {NoStop}%
\bibitem [{\citenamefont {Weisbuch}\ \emph {et~al.}(1992)\citenamefont {Weisbuch}, \citenamefont {Nishioka}, \citenamefont {Ishikawa},\ and\ \citenamefont {Arakawa}}]{art:weisbuch}%
  \BibitemOpen
  \bibfield  {author} {\bibinfo {author} {\bibfnamefont {C.}~\bibnamefont {Weisbuch}}, \bibinfo {author} {\bibfnamefont {M.}~\bibnamefont {Nishioka}}, \bibinfo {author} {\bibfnamefont {A.}~\bibnamefont {Ishikawa}},\ and\ \bibinfo {author} {\bibfnamefont {Y.}~\bibnamefont {Arakawa}},\ }\bibfield  {title} {\bibinfo {title} {Observation of the coupled exciton-photon mode splitting in a semiconductor quantum microcavity},\ }\href {https://doi.org/10.1103/PhysRevLett.69.3314} {\bibfield  {journal} {\bibinfo  {journal} {Phys. Rev. Lett.}\ }\textbf {\bibinfo {volume} {69}},\ \bibinfo {pages} {3314} (\bibinfo {year} {1992})}\BibitemShut {NoStop}%
\bibitem [{\citenamefont {Carusotto}\ and\ \citenamefont {Ciuti}(2013)}]{QuantumFluidOfLight}%
  \BibitemOpen
  \bibfield  {author} {\bibinfo {author} {\bibfnamefont {I.}~\bibnamefont {Carusotto}}\ and\ \bibinfo {author} {\bibfnamefont {C.}~\bibnamefont {Ciuti}},\ }\bibfield  {title} {\bibinfo {title} {Quantum fluids of light},\ }\href {https://doi.org/10.1103/RevModPhys.85.299} {\bibfield  {journal} {\bibinfo  {journal} {Rev. Mod. Phys.}\ }\textbf {\bibinfo {volume} {85}},\ \bibinfo {pages} {299} (\bibinfo {year} {2013})}\BibitemShut {NoStop}%
\bibitem [{\citenamefont {Sanvitto}\ and\ \citenamefont {Kéna-Cohen}(2016)}]{RoadmapPol}%
  \BibitemOpen
  \bibfield  {author} {\bibinfo {author} {\bibfnamefont {D.}~\bibnamefont {Sanvitto}}\ and\ \bibinfo {author} {\bibfnamefont {S.}~\bibnamefont {Kéna-Cohen}},\ }\bibfield  {title} {\bibinfo {title} {The road towards polaritonic devices},\ }\href {https://doi.org/10.1038/nmat4668} {\bibfield  {journal} {\bibinfo  {journal} {Nature Materials}\ }\textbf {\bibinfo {volume} {15}},\ \bibinfo {pages} {1061–1073} (\bibinfo {year} {2016})}\BibitemShut {NoStop}%
\bibitem [{\citenamefont {Gao}\ \emph {et~al.}(2012)\citenamefont {Gao}, \citenamefont {Eldridge}, \citenamefont {Liew}, \citenamefont {Tsintzos}, \citenamefont {Stavrinidis}, \citenamefont {Deligeorgis}, \citenamefont {Hatzopoulos},\ and\ \citenamefont {Savvidis}}]{art:pol_trans}%
  \BibitemOpen
  \bibfield  {author} {\bibinfo {author} {\bibfnamefont {T.}~\bibnamefont {Gao}}, \bibinfo {author} {\bibfnamefont {P.~S.}\ \bibnamefont {Eldridge}}, \bibinfo {author} {\bibfnamefont {T.~C.~H.}\ \bibnamefont {Liew}}, \bibinfo {author} {\bibfnamefont {S.~I.}\ \bibnamefont {Tsintzos}}, \bibinfo {author} {\bibfnamefont {G.}~\bibnamefont {Stavrinidis}}, \bibinfo {author} {\bibfnamefont {G.}~\bibnamefont {Deligeorgis}}, \bibinfo {author} {\bibfnamefont {Z.}~\bibnamefont {Hatzopoulos}},\ and\ \bibinfo {author} {\bibfnamefont {P.~G.}\ \bibnamefont {Savvidis}},\ }\bibfield  {title} {\bibinfo {title} {Polariton condensate transistor switch},\ }\href {https://doi.org/10.1103/PhysRevB.85.235102} {\bibfield  {journal} {\bibinfo  {journal} {Phys. Rev. B}\ }\textbf {\bibinfo {volume} {85}},\ \bibinfo {pages} {235102} (\bibinfo {year} {2012})}\BibitemShut {NoStop}%
\bibitem [{\citenamefont {Ballarini}\ \emph {et~al.}(2013)\citenamefont {Ballarini}, \citenamefont {{De Giorgi}}, \citenamefont {Cancellieri}, \citenamefont {Houdr{\'{e}}}, \citenamefont {Giacobino}, \citenamefont {Cingolani}, \citenamefont {Bramati}, \citenamefont {Gigli},\ and\ \citenamefont {Sanvitto}}]{Ballarini2013a}%
  \BibitemOpen
  \bibfield  {author} {\bibinfo {author} {\bibfnamefont {D.}~\bibnamefont {Ballarini}}, \bibinfo {author} {\bibfnamefont {M.}~\bibnamefont {{De Giorgi}}}, \bibinfo {author} {\bibfnamefont {E.}~\bibnamefont {Cancellieri}}, \bibinfo {author} {\bibfnamefont {R.}~\bibnamefont {Houdr{\'{e}}}}, \bibinfo {author} {\bibfnamefont {E.}~\bibnamefont {Giacobino}}, \bibinfo {author} {\bibfnamefont {R.}~\bibnamefont {Cingolani}}, \bibinfo {author} {\bibfnamefont {A.}~\bibnamefont {Bramati}}, \bibinfo {author} {\bibfnamefont {G.}~\bibnamefont {Gigli}},\ and\ \bibinfo {author} {\bibfnamefont {D.}~\bibnamefont {Sanvitto}},\ }\bibfield  {title} {\bibinfo {title} {{All-optical polariton transistor}},\ }\href {https://doi.org/10.1038/ncomms2734} {\bibfield  {journal} {\bibinfo  {journal} {Nature Communications}\ }\textbf {\bibinfo {volume} {4}},\ \bibinfo {pages} {1778} (\bibinfo {year} {2013})}\BibitemShut {NoStop}%
\bibitem [{\citenamefont {Wu}\ \emph {et~al.}(2021)\citenamefont {Wu}, \citenamefont {Ghosh}, \citenamefont {Su}, \citenamefont {Fieramosca}, \citenamefont {Liew},\ and\ \citenamefont {Xiong}}]{art:non_linear_scatt}%
  \BibitemOpen
  \bibfield  {author} {\bibinfo {author} {\bibfnamefont {J.}~\bibnamefont {Wu}}, \bibinfo {author} {\bibfnamefont {S.}~\bibnamefont {Ghosh}}, \bibinfo {author} {\bibfnamefont {R.}~\bibnamefont {Su}}, \bibinfo {author} {\bibfnamefont {A.}~\bibnamefont {Fieramosca}}, \bibinfo {author} {\bibfnamefont {T.~C.~H.}\ \bibnamefont {Liew}},\ and\ \bibinfo {author} {\bibfnamefont {Q.}~\bibnamefont {Xiong}},\ }\bibfield  {title} {\bibinfo {title} {Nonlinear parametric scattering of exciton polaritons in perovskite microcavities},\ }\href {https://doi.org/10.1021/acs.nanolett.1c00283} {\bibfield  {journal} {\bibinfo  {journal} {Nano Letters}\ }\textbf {\bibinfo {volume} {21}},\ \bibinfo {pages} {3120} (\bibinfo {year} {2021})},\ \Eprint {https://arxiv.org/abs/https://doi.org/10.1021/acs.nanolett.1c00283} {https://doi.org/10.1021/acs.nanolett.1c00283} \BibitemShut {NoStop}%
\bibitem [{\citenamefont {Amo}\ \emph {et~al.}(2011)\citenamefont {Amo}, \citenamefont {Pigeon}, \citenamefont {Sanvitto}, \citenamefont {Sala}, \citenamefont {Hivet}, \citenamefont {Carusotto}, \citenamefont {Pisanello}, \citenamefont {Lem{\'{e}}nager}, \citenamefont {Houdr{\'{e}}}, \citenamefont {Giacobino}, \citenamefont {Ciuti},\ and\ \citenamefont {Bramati}}]{art:amo_sol}%
  \BibitemOpen
  \bibfield  {author} {\bibinfo {author} {\bibfnamefont {A.}~\bibnamefont {Amo}}, \bibinfo {author} {\bibfnamefont {S.}~\bibnamefont {Pigeon}}, \bibinfo {author} {\bibfnamefont {D.}~\bibnamefont {Sanvitto}}, \bibinfo {author} {\bibfnamefont {V.~G.}\ \bibnamefont {Sala}}, \bibinfo {author} {\bibfnamefont {R.}~\bibnamefont {Hivet}}, \bibinfo {author} {\bibfnamefont {I.}~\bibnamefont {Carusotto}}, \bibinfo {author} {\bibfnamefont {F.}~\bibnamefont {Pisanello}}, \bibinfo {author} {\bibfnamefont {G.}~\bibnamefont {Lem{\'{e}}nager}}, \bibinfo {author} {\bibfnamefont {R.}~\bibnamefont {Houdr{\'{e}}}}, \bibinfo {author} {\bibfnamefont {E.}~\bibnamefont {Giacobino}}, \bibinfo {author} {\bibfnamefont {C.}~\bibnamefont {Ciuti}},\ and\ \bibinfo {author} {\bibfnamefont {A.}~\bibnamefont {Bramati}},\ }\bibfield  {title} {\bibinfo {title} {Polariton superfluids reveal quantum hydrodynamic solitons},\ }\href {https://doi.org/10.1126/science.1202307} {\bibfield  {journal} {\bibinfo  {journal} {Science}\ }\textbf {\bibinfo
  {volume} {332}},\ \bibinfo {pages} {1167} (\bibinfo {year} {2011})}\BibitemShut {NoStop}%
\bibitem [{\citenamefont {Nguyen}\ \emph {et~al.}(2013)\citenamefont {Nguyen}, \citenamefont {Vishnevsky}, \citenamefont {Sturm}, \citenamefont {Tanese}, \citenamefont {Solnyshkov}, \citenamefont {Galopin}, \citenamefont {Lema\^{\i}tre}, \citenamefont {Sagnes}, \citenamefont {Amo}, \citenamefont {Malpuech},\ and\ \citenamefont {Bloch}}]{art:hai_tun}%
  \BibitemOpen
  \bibfield  {author} {\bibinfo {author} {\bibfnamefont {H.~S.}\ \bibnamefont {Nguyen}}, \bibinfo {author} {\bibfnamefont {D.}~\bibnamefont {Vishnevsky}}, \bibinfo {author} {\bibfnamefont {C.}~\bibnamefont {Sturm}}, \bibinfo {author} {\bibfnamefont {D.}~\bibnamefont {Tanese}}, \bibinfo {author} {\bibfnamefont {D.}~\bibnamefont {Solnyshkov}}, \bibinfo {author} {\bibfnamefont {E.}~\bibnamefont {Galopin}}, \bibinfo {author} {\bibfnamefont {A.}~\bibnamefont {Lema\^{\i}tre}}, \bibinfo {author} {\bibfnamefont {I.}~\bibnamefont {Sagnes}}, \bibinfo {author} {\bibfnamefont {A.}~\bibnamefont {Amo}}, \bibinfo {author} {\bibfnamefont {G.}~\bibnamefont {Malpuech}},\ and\ \bibinfo {author} {\bibfnamefont {J.}~\bibnamefont {Bloch}},\ }\bibfield  {title} {\bibinfo {title} {Realization of a double-barrier resonant tunneling diode for cavity polaritons},\ }\href {https://doi.org/10.1103/PhysRevLett.110.236601} {\bibfield  {journal} {\bibinfo  {journal} {Phys. Rev. Lett.}\ }\textbf {\bibinfo {volume} {110}},\ \bibinfo {pages}
  {236601} (\bibinfo {year} {2013})}\BibitemShut {NoStop}%
\bibitem [{\citenamefont {Amo}\ \emph {et~al.}(2009)\citenamefont {Amo}, \citenamefont {Lefr{\`{e}}re}, \citenamefont {Pigeon}, \citenamefont {Adrados}, \citenamefont {Ciuti}, \citenamefont {Carusotto}, \citenamefont {Houdr{\'{e}}}, \citenamefont {Giacobino},\ and\ \citenamefont {Bramati}}]{art:superfl}%
  \BibitemOpen
  \bibfield  {author} {\bibinfo {author} {\bibfnamefont {A.}~\bibnamefont {Amo}}, \bibinfo {author} {\bibfnamefont {J.}~\bibnamefont {Lefr{\`{e}}re}}, \bibinfo {author} {\bibfnamefont {S.}~\bibnamefont {Pigeon}}, \bibinfo {author} {\bibfnamefont {C.}~\bibnamefont {Adrados}}, \bibinfo {author} {\bibfnamefont {C.}~\bibnamefont {Ciuti}}, \bibinfo {author} {\bibfnamefont {I.}~\bibnamefont {Carusotto}}, \bibinfo {author} {\bibfnamefont {R.}~\bibnamefont {Houdr{\'{e}}}}, \bibinfo {author} {\bibfnamefont {E.}~\bibnamefont {Giacobino}},\ and\ \bibinfo {author} {\bibfnamefont {A.}~\bibnamefont {Bramati}},\ }\bibfield  {title} {\bibinfo {title} {Superfluidity of polaritons in semiconductor microcavities},\ }\href {https://doi.org/10.1038/nphys1364} {\bibfield  {journal} {\bibinfo  {journal} {Nature Physics}\ }\textbf {\bibinfo {volume} {5}},\ \bibinfo {pages} {805} (\bibinfo {year} {2009})}\BibitemShut {NoStop}%
\bibitem [{\citenamefont {Nguyen}\ \emph {et~al.}(2015)\citenamefont {Nguyen}, \citenamefont {Gerace}, \citenamefont {Carusotto}, \citenamefont {Sanvitto}, \citenamefont {Galopin}, \citenamefont {Lema\^{\i}tre}, \citenamefont {Sagnes}, \citenamefont {Bloch},\ and\ \citenamefont {Amo}}]{art:hai_BH}%
  \BibitemOpen
  \bibfield  {author} {\bibinfo {author} {\bibfnamefont {H.~S.}\ \bibnamefont {Nguyen}}, \bibinfo {author} {\bibfnamefont {D.}~\bibnamefont {Gerace}}, \bibinfo {author} {\bibfnamefont {I.}~\bibnamefont {Carusotto}}, \bibinfo {author} {\bibfnamefont {D.}~\bibnamefont {Sanvitto}}, \bibinfo {author} {\bibfnamefont {E.}~\bibnamefont {Galopin}}, \bibinfo {author} {\bibfnamefont {A.}~\bibnamefont {Lema\^{\i}tre}}, \bibinfo {author} {\bibfnamefont {I.}~\bibnamefont {Sagnes}}, \bibinfo {author} {\bibfnamefont {J.}~\bibnamefont {Bloch}},\ and\ \bibinfo {author} {\bibfnamefont {A.}~\bibnamefont {Amo}},\ }\bibfield  {title} {\bibinfo {title} {Acoustic black hole in a stationary hydrodynamic flow of microcavity polaritons},\ }\href {https://doi.org/10.1103/PhysRevLett.114.036402} {\bibfield  {journal} {\bibinfo  {journal} {Phys. Rev. Lett.}\ }\textbf {\bibinfo {volume} {114}},\ \bibinfo {pages} {036402} (\bibinfo {year} {2015})}\BibitemShut {NoStop}%
\bibitem [{\citenamefont {Liran}\ \emph {et~al.}(2018)\citenamefont {Liran}, \citenamefont {Rosenberg}, \citenamefont {West}, \citenamefont {Pfeiffer},\ and\ \citenamefont {Rapaport}}]{art:full_guided}%
  \BibitemOpen
  \bibfield  {author} {\bibinfo {author} {\bibfnamefont {D.}~\bibnamefont {Liran}}, \bibinfo {author} {\bibfnamefont {I.}~\bibnamefont {Rosenberg}}, \bibinfo {author} {\bibfnamefont {K.}~\bibnamefont {West}}, \bibinfo {author} {\bibfnamefont {L.}~\bibnamefont {Pfeiffer}},\ and\ \bibinfo {author} {\bibfnamefont {R.}~\bibnamefont {Rapaport}},\ }\bibfield  {title} {\bibinfo {title} {Fully guided electrically controlled exciton polaritons},\ }\href@noop {} {\bibfield  {journal} {\bibinfo  {journal} {ACS Photonics}\ }\textbf {\bibinfo {volume} {5}},\ \bibinfo {pages} {4249} (\bibinfo {year} {2018})}\BibitemShut {NoStop}%
\bibitem [{\citenamefont {Wertz}\ \emph {et~al.}(2010)\citenamefont {Wertz}, \citenamefont {Ferrier}, \citenamefont {Solnyshkov}, \citenamefont {Johne}, \citenamefont {Sanvitto}, \citenamefont {Lema{\^{\i}}tre}, \citenamefont {Sagnes}, \citenamefont {Grousson}, \citenamefont {Kavokin}, \citenamefont {Senellart}, \citenamefont {Malpuech},\ and\ \citenamefont {Bloch}}]{art:wertz}%
  \BibitemOpen
  \bibfield  {author} {\bibinfo {author} {\bibfnamefont {E.}~\bibnamefont {Wertz}}, \bibinfo {author} {\bibfnamefont {L.}~\bibnamefont {Ferrier}}, \bibinfo {author} {\bibfnamefont {D.~D.}\ \bibnamefont {Solnyshkov}}, \bibinfo {author} {\bibfnamefont {R.}~\bibnamefont {Johne}}, \bibinfo {author} {\bibfnamefont {D.}~\bibnamefont {Sanvitto}}, \bibinfo {author} {\bibfnamefont {A.}~\bibnamefont {Lema{\^{\i}}tre}}, \bibinfo {author} {\bibfnamefont {I.}~\bibnamefont {Sagnes}}, \bibinfo {author} {\bibfnamefont {R.}~\bibnamefont {Grousson}}, \bibinfo {author} {\bibfnamefont {A.~V.}\ \bibnamefont {Kavokin}}, \bibinfo {author} {\bibfnamefont {P.}~\bibnamefont {Senellart}}, \bibinfo {author} {\bibfnamefont {G.}~\bibnamefont {Malpuech}},\ and\ \bibinfo {author} {\bibfnamefont {J.}~\bibnamefont {Bloch}},\ }\bibfield  {title} {\bibinfo {title} {Spontaneous formation and optical manipulation of extended polariton condensates},\ }\href {https://doi.org/10.1038/nphys1750} {\bibfield  {journal} {\bibinfo  {journal} {Nature
  Physics}\ }\textbf {\bibinfo {volume} {6}},\ \bibinfo {pages} {860} (\bibinfo {year} {2010})}\BibitemShut {NoStop}%
\bibitem [{\citenamefont {Steger}\ \emph {et~al.}(2013)\citenamefont {Steger}, \citenamefont {Liu}, \citenamefont {Nelsen}, \citenamefont {Gautham}, \citenamefont {Snoke}, \citenamefont {Balili}, \citenamefont {Pfeiffer},\ and\ \citenamefont {West}}]{art:steger}%
  \BibitemOpen
  \bibfield  {author} {\bibinfo {author} {\bibfnamefont {M.}~\bibnamefont {Steger}}, \bibinfo {author} {\bibfnamefont {G.}~\bibnamefont {Liu}}, \bibinfo {author} {\bibfnamefont {B.}~\bibnamefont {Nelsen}}, \bibinfo {author} {\bibfnamefont {C.}~\bibnamefont {Gautham}}, \bibinfo {author} {\bibfnamefont {D.~W.}\ \bibnamefont {Snoke}}, \bibinfo {author} {\bibfnamefont {R.}~\bibnamefont {Balili}}, \bibinfo {author} {\bibfnamefont {L.}~\bibnamefont {Pfeiffer}},\ and\ \bibinfo {author} {\bibfnamefont {K.}~\bibnamefont {West}},\ }\bibfield  {title} {\bibinfo {title} {Long-range ballistic motion and coherent flow of long-lifetime polaritons},\ }\href {https://doi.org/10.1103/PhysRevB.88.235314} {\bibfield  {journal} {\bibinfo  {journal} {Phys. Rev. B}\ }\textbf {\bibinfo {volume} {88}},\ \bibinfo {pages} {235314} (\bibinfo {year} {2013})}\BibitemShut {NoStop}%
\bibitem [{\citenamefont {Freixanet}\ \emph {et~al.}(2000)\citenamefont {Freixanet}, \citenamefont {Sermage}, \citenamefont {Tiberj},\ and\ \citenamefont {Planel}}]{art:freixanet}%
  \BibitemOpen
  \bibfield  {author} {\bibinfo {author} {\bibfnamefont {T.}~\bibnamefont {Freixanet}}, \bibinfo {author} {\bibfnamefont {B.}~\bibnamefont {Sermage}}, \bibinfo {author} {\bibfnamefont {A.}~\bibnamefont {Tiberj}},\ and\ \bibinfo {author} {\bibfnamefont {R.}~\bibnamefont {Planel}},\ }\bibfield  {title} {\bibinfo {title} {In-plane propagation of excitonic cavity polaritons},\ }\href {https://doi.org/10.1103/PhysRevB.61.7233} {\bibfield  {journal} {\bibinfo  {journal} {Phys. Rev. B}\ }\textbf {\bibinfo {volume} {61}},\ \bibinfo {pages} {7233} (\bibinfo {year} {2000})}\BibitemShut {NoStop}%
\bibitem [{\citenamefont {Langbein}\ \emph {et~al.}(2007)\citenamefont {Langbein}, \citenamefont {Shelykh}, \citenamefont {Solnyshkov}, \citenamefont {Malpuech}, \citenamefont {Rubo},\ and\ \citenamefont {Kavokin}}]{art:langbein}%
  \BibitemOpen
  \bibfield  {author} {\bibinfo {author} {\bibfnamefont {W.}~\bibnamefont {Langbein}}, \bibinfo {author} {\bibfnamefont {I.}~\bibnamefont {Shelykh}}, \bibinfo {author} {\bibfnamefont {D.}~\bibnamefont {Solnyshkov}}, \bibinfo {author} {\bibfnamefont {G.}~\bibnamefont {Malpuech}}, \bibinfo {author} {\bibfnamefont {Y.}~\bibnamefont {Rubo}},\ and\ \bibinfo {author} {\bibfnamefont {A.}~\bibnamefont {Kavokin}},\ }\bibfield  {title} {\bibinfo {title} {Polarization beats in ballistic propagation of exciton-polaritons in microcavities},\ }\href {https://doi.org/10.1103/PhysRevB.75.075323} {\bibfield  {journal} {\bibinfo  {journal} {Phys. Rev. B}\ }\textbf {\bibinfo {volume} {75}},\ \bibinfo {pages} {075323} (\bibinfo {year} {2007})}\BibitemShut {NoStop}%
\bibitem [{\citenamefont {Su{\'{a}}rez-Forero}\ \emph {et~al.}(2020)\citenamefont {Su{\'{a}}rez-Forero}, \citenamefont {Ardizzone}, \citenamefont {da~Silva}, \citenamefont {Reindl}, \citenamefont {Fieramosca}, \citenamefont {Polimeno}, \citenamefont {Giorgi}, \citenamefont {Dominici}, \citenamefont {Pfeiffer}, \citenamefont {Gigli}, \citenamefont {Ballarini}, \citenamefont {Laussy}, \citenamefont {Rastelli},\ and\ \citenamefont {Sanvitto}}]{art:surez}%
  \BibitemOpen
  \bibfield  {author} {\bibinfo {author} {\bibfnamefont {D.~G.}\ \bibnamefont {Su{\'{a}}rez-Forero}}, \bibinfo {author} {\bibfnamefont {V.}~\bibnamefont {Ardizzone}}, \bibinfo {author} {\bibfnamefont {S.~F.~C.}\ \bibnamefont {da~Silva}}, \bibinfo {author} {\bibfnamefont {M.}~\bibnamefont {Reindl}}, \bibinfo {author} {\bibfnamefont {A.}~\bibnamefont {Fieramosca}}, \bibinfo {author} {\bibfnamefont {L.}~\bibnamefont {Polimeno}}, \bibinfo {author} {\bibfnamefont {M.~D.}\ \bibnamefont {Giorgi}}, \bibinfo {author} {\bibfnamefont {L.}~\bibnamefont {Dominici}}, \bibinfo {author} {\bibfnamefont {L.~N.}\ \bibnamefont {Pfeiffer}}, \bibinfo {author} {\bibfnamefont {G.}~\bibnamefont {Gigli}}, \bibinfo {author} {\bibfnamefont {D.}~\bibnamefont {Ballarini}}, \bibinfo {author} {\bibfnamefont {F.}~\bibnamefont {Laussy}}, \bibinfo {author} {\bibfnamefont {A.}~\bibnamefont {Rastelli}},\ and\ \bibinfo {author} {\bibfnamefont {D.}~\bibnamefont {Sanvitto}},\ }\bibfield  {title} {\bibinfo {title} {Quantum hydrodynamics of a single
  particle},\ }\href {https://doi.org/10.1038/s41377-020-0324-x} {\bibfield  {journal} {\bibinfo  {journal} {Light: Science $\&$ Applications}\ }\textbf {\bibinfo {volume} {9}},\ \bibinfo {pages} {85} (\bibinfo {year} {2020})}\BibitemShut {NoStop}%
\bibitem [{\citenamefont {Walker}\ \emph {et~al.}(2013)\citenamefont {Walker}, \citenamefont {Tinkler}, \citenamefont {Durska}, \citenamefont {Whittaker}, \citenamefont {Luxmoore}, \citenamefont {Royall}, \citenamefont {Krizhanovskii}, \citenamefont {Skolnick}, \citenamefont {Farrer},\ and\ \citenamefont {Ritchie}}]{art:Walker}%
  \BibitemOpen
  \bibfield  {author} {\bibinfo {author} {\bibfnamefont {P.~M.}\ \bibnamefont {Walker}}, \bibinfo {author} {\bibfnamefont {L.}~\bibnamefont {Tinkler}}, \bibinfo {author} {\bibfnamefont {M.}~\bibnamefont {Durska}}, \bibinfo {author} {\bibfnamefont {D.~M.}\ \bibnamefont {Whittaker}}, \bibinfo {author} {\bibfnamefont {I.~J.}\ \bibnamefont {Luxmoore}}, \bibinfo {author} {\bibfnamefont {B.}~\bibnamefont {Royall}}, \bibinfo {author} {\bibfnamefont {D.~N.}\ \bibnamefont {Krizhanovskii}}, \bibinfo {author} {\bibfnamefont {M.~S.}\ \bibnamefont {Skolnick}}, \bibinfo {author} {\bibfnamefont {I.}~\bibnamefont {Farrer}},\ and\ \bibinfo {author} {\bibfnamefont {D.~A.}\ \bibnamefont {Ritchie}},\ }\bibfield  {title} {\bibinfo {title} {Exciton polaritons in semiconductor waveguides},\ }\href {https://doi.org/10.1063/1.4773590} {\bibfield  {journal} {\bibinfo  {journal} {Applied Physics Letters}\ }\textbf {\bibinfo {volume} {102}},\ \bibinfo {pages} {012109} (\bibinfo {year} {2013})}\BibitemShut {NoStop}%
\bibitem [{\citenamefont {Tanese}\ \emph {et~al.}(2012)\citenamefont {Tanese}, \citenamefont {Solnyshkov}, \citenamefont {Amo}, \citenamefont {Ferrier}, \citenamefont {Bernet-Rollande}, \citenamefont {Wertz}, \citenamefont {Sagnes}, \citenamefont {Lema\^{\i}tre}, \citenamefont {Senellart}, \citenamefont {Malpuech},\ and\ \citenamefont {Bloch}}]{Tanese2012}%
  \BibitemOpen
  \bibfield  {author} {\bibinfo {author} {\bibfnamefont {D.}~\bibnamefont {Tanese}}, \bibinfo {author} {\bibfnamefont {D.~D.}\ \bibnamefont {Solnyshkov}}, \bibinfo {author} {\bibfnamefont {A.}~\bibnamefont {Amo}}, \bibinfo {author} {\bibfnamefont {L.}~\bibnamefont {Ferrier}}, \bibinfo {author} {\bibfnamefont {E.}~\bibnamefont {Bernet-Rollande}}, \bibinfo {author} {\bibfnamefont {E.}~\bibnamefont {Wertz}}, \bibinfo {author} {\bibfnamefont {I.}~\bibnamefont {Sagnes}}, \bibinfo {author} {\bibfnamefont {A.}~\bibnamefont {Lema\^{\i}tre}}, \bibinfo {author} {\bibfnamefont {P.}~\bibnamefont {Senellart}}, \bibinfo {author} {\bibfnamefont {G.}~\bibnamefont {Malpuech}},\ and\ \bibinfo {author} {\bibfnamefont {J.}~\bibnamefont {Bloch}},\ }\bibfield  {title} {\bibinfo {title} {Backscattering suppression in supersonic 1d polariton condensates},\ }\href {https://doi.org/10.1103/PhysRevLett.108.036405} {\bibfield  {journal} {\bibinfo  {journal} {Phys. Rev. Lett.}\ }\textbf {\bibinfo {volume} {108}},\ \bibinfo {pages}
  {036405} (\bibinfo {year} {2012})}\BibitemShut {NoStop}%
\bibitem [{\citenamefont {Marsault}\ \emph {et~al.}(2015)\citenamefont {Marsault}, \citenamefont {Nguyen}, \citenamefont {Tanese}, \citenamefont {Lema{\^{\i}}tre}, \citenamefont {Galopin}, \citenamefont {Sagnes}, \citenamefont {Amo},\ and\ \citenamefont {Bloch}}]{art:Marsault}%
  \BibitemOpen
  \bibfield  {author} {\bibinfo {author} {\bibfnamefont {F.}~\bibnamefont {Marsault}}, \bibinfo {author} {\bibfnamefont {H.~S.}\ \bibnamefont {Nguyen}}, \bibinfo {author} {\bibfnamefont {D.}~\bibnamefont {Tanese}}, \bibinfo {author} {\bibfnamefont {A.}~\bibnamefont {Lema{\^{\i}}tre}}, \bibinfo {author} {\bibfnamefont {E.}~\bibnamefont {Galopin}}, \bibinfo {author} {\bibfnamefont {I.}~\bibnamefont {Sagnes}}, \bibinfo {author} {\bibfnamefont {A.}~\bibnamefont {Amo}},\ and\ \bibinfo {author} {\bibfnamefont {J.}~\bibnamefont {Bloch}},\ }\bibfield  {title} {\bibinfo {title} {Realization of an all optical exciton-polariton router},\ }\href {https://doi.org/10.1063/1.4936158} {\bibfield  {journal} {\bibinfo  {journal} {Applied Physics Letters}\ }\textbf {\bibinfo {volume} {107}},\ \bibinfo {pages} {201115} (\bibinfo {year} {2015})}\BibitemShut {NoStop}%
\bibitem [{\citenamefont {Franke}\ \emph {et~al.}(2012)\citenamefont {Franke}, \citenamefont {Sturm}, \citenamefont {Schmidt-Grund}, \citenamefont {Wagner},\ and\ \citenamefont {Grundmann}}]{art:Franke}%
  \BibitemOpen
  \bibfield  {author} {\bibinfo {author} {\bibfnamefont {H.}~\bibnamefont {Franke}}, \bibinfo {author} {\bibfnamefont {C.}~\bibnamefont {Sturm}}, \bibinfo {author} {\bibfnamefont {R.}~\bibnamefont {Schmidt-Grund}}, \bibinfo {author} {\bibfnamefont {G.}~\bibnamefont {Wagner}},\ and\ \bibinfo {author} {\bibfnamefont {M.}~\bibnamefont {Grundmann}},\ }\bibfield  {title} {\bibinfo {title} {Ballistic propagation of exciton{\textendash}polariton condensates in a {ZnO}-based microcavity},\ }\href {https://doi.org/10.1088/1367-2630/14/1/013037} {\bibfield  {journal} {\bibinfo  {journal} {New Journal of Physics}\ }\textbf {\bibinfo {volume} {14}},\ \bibinfo {pages} {013037} (\bibinfo {year} {2012})}\BibitemShut {NoStop}%
\bibitem [{\citenamefont {Ghosh}\ \emph {et~al.}(2022)\citenamefont {Ghosh}, \citenamefont {Su}, \citenamefont {Zhao}, \citenamefont {Fieramosca}, \citenamefont {Wu}, \citenamefont {Li}, \citenamefont {Zhang}, \citenamefont {Li}, \citenamefont {Chen}, \citenamefont {Liew}, \citenamefont {Sanvitto},\ and\ \citenamefont {Xiong}}]{art:Ghosh_ucavityRT}%
  \BibitemOpen
  \bibfield  {author} {\bibinfo {author} {\bibfnamefont {S.}~\bibnamefont {Ghosh}}, \bibinfo {author} {\bibfnamefont {R.}~\bibnamefont {Su}}, \bibinfo {author} {\bibfnamefont {J.}~\bibnamefont {Zhao}}, \bibinfo {author} {\bibfnamefont {A.}~\bibnamefont {Fieramosca}}, \bibinfo {author} {\bibfnamefont {J.}~\bibnamefont {Wu}}, \bibinfo {author} {\bibfnamefont {T.}~\bibnamefont {Li}}, \bibinfo {author} {\bibfnamefont {Q.}~\bibnamefont {Zhang}}, \bibinfo {author} {\bibfnamefont {F.}~\bibnamefont {Li}}, \bibinfo {author} {\bibfnamefont {Z.}~\bibnamefont {Chen}}, \bibinfo {author} {\bibfnamefont {T.}~\bibnamefont {Liew}}, \bibinfo {author} {\bibfnamefont {D.}~\bibnamefont {Sanvitto}},\ and\ \bibinfo {author} {\bibfnamefont {Q.}~\bibnamefont {Xiong}},\ }\bibfield  {title} {\bibinfo {title} {Microcavity exciton polaritons at room temperature},\ }\href {https://doi.org/10.3788/pi.2022.r04} {\bibfield  {journal} {\bibinfo  {journal} {Photonics Insights}\ }\textbf {\bibinfo {volume} {1}},\ \bibinfo {pages} {R04}
  (\bibinfo {year} {2022})}\BibitemShut {NoStop}%
\bibitem [{\citenamefont {Lerario}\ \emph {et~al.}(2016)\citenamefont {Lerario}, \citenamefont {Ballarini}, \citenamefont {Fieramosca}, \citenamefont {Cannavale}, \citenamefont {Genco}, \citenamefont {Mangione}, \citenamefont {Gambino}, \citenamefont {Dominici}, \citenamefont {Giorgi}, \citenamefont {Gigli},\ and\ \citenamefont {Sanvitto}}]{art:high_speed}%
  \BibitemOpen
  \bibfield  {author} {\bibinfo {author} {\bibfnamefont {G.}~\bibnamefont {Lerario}}, \bibinfo {author} {\bibfnamefont {D.}~\bibnamefont {Ballarini}}, \bibinfo {author} {\bibfnamefont {A.}~\bibnamefont {Fieramosca}}, \bibinfo {author} {\bibfnamefont {A.}~\bibnamefont {Cannavale}}, \bibinfo {author} {\bibfnamefont {A.}~\bibnamefont {Genco}}, \bibinfo {author} {\bibfnamefont {F.}~\bibnamefont {Mangione}}, \bibinfo {author} {\bibfnamefont {S.}~\bibnamefont {Gambino}}, \bibinfo {author} {\bibfnamefont {L.}~\bibnamefont {Dominici}}, \bibinfo {author} {\bibfnamefont {M.~D.}\ \bibnamefont {Giorgi}}, \bibinfo {author} {\bibfnamefont {G.}~\bibnamefont {Gigli}},\ and\ \bibinfo {author} {\bibfnamefont {D.}~\bibnamefont {Sanvitto}},\ }\bibfield  {title} {\bibinfo {title} {High-speed flow of interacting organic polaritons},\ }\href {https://doi.org/10.1038/lsa.2016.212} {\bibfield  {journal} {\bibinfo  {journal} {Light: Science $\&$ Applications}\ }\textbf {\bibinfo {volume} {6}},\ \bibinfo {pages} {e16212} (\bibinfo
  {year} {2016})}\BibitemShut {NoStop}%
\bibitem [{\citenamefont {Hou}\ \emph {et~al.}(2020{\natexlab{a}})\citenamefont {Hou}, \citenamefont {Khatoniar}, \citenamefont {Ding}, \citenamefont {Qu}, \citenamefont {Napolov}, \citenamefont {Menon},\ and\ \citenamefont {Forrest}}]{arT:Hou}%
  \BibitemOpen
  \bibfield  {author} {\bibinfo {author} {\bibfnamefont {S.}~\bibnamefont {Hou}}, \bibinfo {author} {\bibfnamefont {M.}~\bibnamefont {Khatoniar}}, \bibinfo {author} {\bibfnamefont {K.}~\bibnamefont {Ding}}, \bibinfo {author} {\bibfnamefont {Y.}~\bibnamefont {Qu}}, \bibinfo {author} {\bibfnamefont {A.}~\bibnamefont {Napolov}}, \bibinfo {author} {\bibfnamefont {V.~M.}\ \bibnamefont {Menon}},\ and\ \bibinfo {author} {\bibfnamefont {S.~R.}\ \bibnamefont {Forrest}},\ }\bibfield  {title} {\bibinfo {title} {Ultralong-range energy transport in a disordered organic semiconductor at room temperature via coherent exciton-polariton propagation},\ }\href {https://doi.org/10.1002/adma.202002127} {\bibfield  {journal} {\bibinfo  {journal} {Advanced Materials}\ }\textbf {\bibinfo {volume} {32}},\ \bibinfo {pages} {2002127} (\bibinfo {year} {2020}{\natexlab{a}})}\BibitemShut {NoStop}%
\bibitem [{\citenamefont {Zhang}\ \emph {et~al.}(2011)\citenamefont {Zhang}, \citenamefont {Zou}, \citenamefont {Yan}, \citenamefont {Hao}, \citenamefont {Sun}, \citenamefont {Han}, \citenamefont {Zhao},\ and\ \citenamefont {Yao}}]{art:Zhang}%
  \BibitemOpen
  \bibfield  {author} {\bibinfo {author} {\bibfnamefont {C.}~\bibnamefont {Zhang}}, \bibinfo {author} {\bibfnamefont {C.-L.}\ \bibnamefont {Zou}}, \bibinfo {author} {\bibfnamefont {Y.}~\bibnamefont {Yan}}, \bibinfo {author} {\bibfnamefont {R.}~\bibnamefont {Hao}}, \bibinfo {author} {\bibfnamefont {F.-W.}\ \bibnamefont {Sun}}, \bibinfo {author} {\bibfnamefont {Z.-F.}\ \bibnamefont {Han}}, \bibinfo {author} {\bibfnamefont {Y.~S.}\ \bibnamefont {Zhao}},\ and\ \bibinfo {author} {\bibfnamefont {J.}~\bibnamefont {Yao}},\ }\bibfield  {title} {\bibinfo {title} {Two-photon pumped lasing in single-crystal organic nanowire exciton polariton resonators},\ }\href {https://doi.org/10.1021/ja200549v} {\bibfield  {journal} {\bibinfo  {journal} {Journal of the American Chemical Society}\ }\textbf {\bibinfo {volume} {133}},\ \bibinfo {pages} {7276} (\bibinfo {year} {2011})}\BibitemShut {NoStop}%
\bibitem [{\citenamefont {Ciers}\ \emph {et~al.}(2017)\citenamefont {Ciers}, \citenamefont {Roch}, \citenamefont {Carlin}, \citenamefont {Jacopin}, \citenamefont {Butt\'e},\ and\ \citenamefont {Grandjean}}]{art:Ciers}%
  \BibitemOpen
  \bibfield  {author} {\bibinfo {author} {\bibfnamefont {J.}~\bibnamefont {Ciers}}, \bibinfo {author} {\bibfnamefont {J.~G.}\ \bibnamefont {Roch}}, \bibinfo {author} {\bibfnamefont {J.-F.}\ \bibnamefont {Carlin}}, \bibinfo {author} {\bibfnamefont {G.}~\bibnamefont {Jacopin}}, \bibinfo {author} {\bibfnamefont {R.}~\bibnamefont {Butt\'e}},\ and\ \bibinfo {author} {\bibfnamefont {N.}~\bibnamefont {Grandjean}},\ }\bibfield  {title} {\bibinfo {title} {Propagating polaritons in iii-nitride slab waveguides},\ }\href {https://doi.org/10.1103/PhysRevApplied.7.034019} {\bibfield  {journal} {\bibinfo  {journal} {Phys. Rev. Appl.}\ }\textbf {\bibinfo {volume} {7}},\ \bibinfo {pages} {034019} (\bibinfo {year} {2017})}\BibitemShut {NoStop}%
\bibitem [{\citenamefont {Barachati}\ \emph {et~al.}(2018)\citenamefont {Barachati}, \citenamefont {Fieramosca}, \citenamefont {Hafezian}, \citenamefont {Gu}, \citenamefont {Chakraborty}, \citenamefont {Ballarini}, \citenamefont {Martinu}, \citenamefont {Menon}, \citenamefont {Sanvitto},\ and\ \citenamefont {K{\'{e}}na-Cohen}}]{art:barachati}%
  \BibitemOpen
  \bibfield  {author} {\bibinfo {author} {\bibfnamefont {F.}~\bibnamefont {Barachati}}, \bibinfo {author} {\bibfnamefont {A.}~\bibnamefont {Fieramosca}}, \bibinfo {author} {\bibfnamefont {S.}~\bibnamefont {Hafezian}}, \bibinfo {author} {\bibfnamefont {J.}~\bibnamefont {Gu}}, \bibinfo {author} {\bibfnamefont {B.}~\bibnamefont {Chakraborty}}, \bibinfo {author} {\bibfnamefont {D.}~\bibnamefont {Ballarini}}, \bibinfo {author} {\bibfnamefont {L.}~\bibnamefont {Martinu}}, \bibinfo {author} {\bibfnamefont {V.}~\bibnamefont {Menon}}, \bibinfo {author} {\bibfnamefont {D.}~\bibnamefont {Sanvitto}},\ and\ \bibinfo {author} {\bibfnamefont {S.}~\bibnamefont {K{\'{e}}na-Cohen}},\ }\bibfield  {title} {\bibinfo {title} {Interacting polariton fluids in a monolayer of tungsten disulfide},\ }\href {https://doi.org/10.1038/s41565-018-0219-7} {\bibfield  {journal} {\bibinfo  {journal} {Nature Nanotechnology}\ }\textbf {\bibinfo {volume} {13}},\ \bibinfo {pages} {906} (\bibinfo {year} {2018})}\BibitemShut {NoStop}%
\bibitem [{\citenamefont {Liu}\ \emph {et~al.}(2023)\citenamefont {Liu}, \citenamefont {Lynch}, \citenamefont {Zhao}, \citenamefont {Conran}, \citenamefont {McAleese}, \citenamefont {Jariwala},\ and\ \citenamefont {Forrest}}]{Liu2023}%
  \BibitemOpen
  \bibfield  {author} {\bibinfo {author} {\bibfnamefont {B.}~\bibnamefont {Liu}}, \bibinfo {author} {\bibfnamefont {J.}~\bibnamefont {Lynch}}, \bibinfo {author} {\bibfnamefont {H.}~\bibnamefont {Zhao}}, \bibinfo {author} {\bibfnamefont {B.~R.}\ \bibnamefont {Conran}}, \bibinfo {author} {\bibfnamefont {C.}~\bibnamefont {McAleese}}, \bibinfo {author} {\bibfnamefont {D.}~\bibnamefont {Jariwala}},\ and\ \bibinfo {author} {\bibfnamefont {S.~R.}\ \bibnamefont {Forrest}},\ }\bibfield  {title} {\bibinfo {title} {Long-range propagation of exciton-polaritons in large-area 2d semiconductor monolayers},\ }\href {https://doi.org/10.1021/acsnano.3c03485} {\bibfield  {journal} {\bibinfo  {journal} {ACS Nano}\ }\textbf {\bibinfo {volume} {17}},\ \bibinfo {pages} {14442} (\bibinfo {year} {2023})},\ \bibinfo {note} {pMID: 37489978},\ \Eprint {https://arxiv.org/abs/https://doi.org/10.1021/acsnano.3c03485} {https://doi.org/10.1021/acsnano.3c03485} \BibitemShut {NoStop}%
\bibitem [{\citenamefont {Wurdack}\ \emph {et~al.}(2022)\citenamefont {Wurdack}, \citenamefont {Estrecho}, \citenamefont {Todd}, \citenamefont {Yun}, \citenamefont {Pieczarka}, \citenamefont {Earl}, \citenamefont {Davis}, \citenamefont {Schneider}, \citenamefont {Truscott},\ and\ \citenamefont {Ostrovskaya}}]{art:Wurdak_prop}%
  \BibitemOpen
  \bibfield  {author} {\bibinfo {author} {\bibfnamefont {M.}~\bibnamefont {Wurdack}}, \bibinfo {author} {\bibfnamefont {E.}~\bibnamefont {Estrecho}}, \bibinfo {author} {\bibfnamefont {S.}~\bibnamefont {Todd}}, \bibinfo {author} {\bibfnamefont {T.}~\bibnamefont {Yun}}, \bibinfo {author} {\bibfnamefont {M.}~\bibnamefont {Pieczarka}}, \bibinfo {author} {\bibfnamefont {S.~K.}\ \bibnamefont {Earl}}, \bibinfo {author} {\bibfnamefont {J.~A.}\ \bibnamefont {Davis}}, \bibinfo {author} {\bibfnamefont {C.}~\bibnamefont {Schneider}}, \bibinfo {author} {\bibfnamefont {A.~G.}\ \bibnamefont {Truscott}},\ and\ \bibinfo {author} {\bibfnamefont {E.~A.}\ \bibnamefont {Ostrovskaya}},\ }\bibfield  {title} {\bibinfo {title} {Author correction: Motional narrowing, ballistic transport, and trapping of room-temperature exciton polaritons in an atomically-thin semiconductor},\ }\href@noop {} {\bibfield  {journal} {\bibinfo  {journal} {Nat. Commun.}\ }\textbf {\bibinfo {volume} {13}},\ \bibinfo {pages} {2636} (\bibinfo {year}
  {2022})}\BibitemShut {NoStop}%
\bibitem [{\citenamefont {Su}\ \emph {et~al.}(2018)\citenamefont {Su}, \citenamefont {Wang}, \citenamefont {Zhao}, \citenamefont {Xing}, \citenamefont {Zhao}, \citenamefont {Diederichs}, \citenamefont {Liew},\ and\ \citenamefont {Xiong}}]{art:su}%
  \BibitemOpen
  \bibfield  {author} {\bibinfo {author} {\bibfnamefont {R.}~\bibnamefont {Su}}, \bibinfo {author} {\bibfnamefont {J.}~\bibnamefont {Wang}}, \bibinfo {author} {\bibfnamefont {J.}~\bibnamefont {Zhao}}, \bibinfo {author} {\bibfnamefont {J.}~\bibnamefont {Xing}}, \bibinfo {author} {\bibfnamefont {W.}~\bibnamefont {Zhao}}, \bibinfo {author} {\bibfnamefont {C.}~\bibnamefont {Diederichs}}, \bibinfo {author} {\bibfnamefont {T.~C.~H.}\ \bibnamefont {Liew}},\ and\ \bibinfo {author} {\bibfnamefont {Q.}~\bibnamefont {Xiong}},\ }\bibfield  {title} {\bibinfo {title} {Room temperature long-range coherent exciton polariton condensate flow in lead halide perovskites},\ }\href {https://doi.org/10.1126/sciadv.aau0244} {\bibfield  {journal} {\bibinfo  {journal} {Science Advances}\ }\textbf {\bibinfo {volume} {4}},\ \bibinfo {pages} {eaau0244} (\bibinfo {year} {2018})},\ \Eprint {https://arxiv.org/abs/https://www.science.org/doi/pdf/10.1126/sciadv.aau0244} {https://www.science.org/doi/pdf/10.1126/sciadv.aau0244} \BibitemShut
  {NoStop}%
\bibitem [{\citenamefont {Peng}\ \emph {et~al.}(2022{\natexlab{a}})\citenamefont {Peng}, \citenamefont {Tao}, \citenamefont {Haeberl{\'{e}}}, \citenamefont {Li}, \citenamefont {Jin}, \citenamefont {Fleming}, \citenamefont {K{\'{e}}na-Cohen}, \citenamefont {Zhang},\ and\ \citenamefont {Bao}}]{Peng2022}%
  \BibitemOpen
  \bibfield  {author} {\bibinfo {author} {\bibfnamefont {K.}~\bibnamefont {Peng}}, \bibinfo {author} {\bibfnamefont {R.}~\bibnamefont {Tao}}, \bibinfo {author} {\bibfnamefont {L.}~\bibnamefont {Haeberl{\'{e}}}}, \bibinfo {author} {\bibfnamefont {Q.}~\bibnamefont {Li}}, \bibinfo {author} {\bibfnamefont {D.}~\bibnamefont {Jin}}, \bibinfo {author} {\bibfnamefont {G.~R.}\ \bibnamefont {Fleming}}, \bibinfo {author} {\bibfnamefont {S.}~\bibnamefont {K{\'{e}}na-Cohen}}, \bibinfo {author} {\bibfnamefont {X.}~\bibnamefont {Zhang}},\ and\ \bibinfo {author} {\bibfnamefont {W.}~\bibnamefont {Bao}},\ }\bibfield  {title} {\bibinfo {title} {{Room-temperature polariton quantum fluids in halide perovskites}},\ }\href {https://doi.org/10.1038/s41467-022-34987-y} {\bibfield  {journal} {\bibinfo  {journal} {Nature Communications}\ }\textbf {\bibinfo {volume} {13}},\ \bibinfo {pages} {7388} (\bibinfo {year} {2022}{\natexlab{a}})}\BibitemShut {NoStop}%
\bibitem [{\citenamefont {Xu}\ \emph {et~al.}(2023)\citenamefont {Xu}, \citenamefont {Mandal}, \citenamefont {Baxter}, \citenamefont {Cheng}, \citenamefont {Lee}, \citenamefont {Su}, \citenamefont {Liu}, \citenamefont {Reichman},\ and\ \citenamefont {Delor}}]{Xu2023}%
  \BibitemOpen
  \bibfield  {author} {\bibinfo {author} {\bibfnamefont {D.}~\bibnamefont {Xu}}, \bibinfo {author} {\bibfnamefont {A.}~\bibnamefont {Mandal}}, \bibinfo {author} {\bibfnamefont {J.~M.}\ \bibnamefont {Baxter}}, \bibinfo {author} {\bibfnamefont {S.-W.}\ \bibnamefont {Cheng}}, \bibinfo {author} {\bibfnamefont {I.}~\bibnamefont {Lee}}, \bibinfo {author} {\bibfnamefont {H.}~\bibnamefont {Su}}, \bibinfo {author} {\bibfnamefont {S.}~\bibnamefont {Liu}}, \bibinfo {author} {\bibfnamefont {D.~R.}\ \bibnamefont {Reichman}},\ and\ \bibinfo {author} {\bibfnamefont {M.}~\bibnamefont {Delor}},\ }\bibfield  {title} {\bibinfo {title} {{Ultrafast imaging of polariton propagation and interactions}},\ }\href {https://doi.org/10.1038/s41467-023-39550-x} {\bibfield  {journal} {\bibinfo  {journal} {Nature Communications}\ }\textbf {\bibinfo {volume} {14}},\ \bibinfo {pages} {3881} (\bibinfo {year} {2023})}\BibitemShut {NoStop}%
\bibitem [{\citenamefont {Chen}\ \emph {et~al.}(2023)\citenamefont {Chen}, \citenamefont {Shi}, \citenamefont {Gan}, \citenamefont {Liu}, \citenamefont {Li}, \citenamefont {Ghosh},\ and\ \citenamefont {Xiong}}]{Chen2023}%
  \BibitemOpen
  \bibfield  {author} {\bibinfo {author} {\bibfnamefont {Y.}~\bibnamefont {Chen}}, \bibinfo {author} {\bibfnamefont {Y.}~\bibnamefont {Shi}}, \bibinfo {author} {\bibfnamefont {Y.}~\bibnamefont {Gan}}, \bibinfo {author} {\bibfnamefont {H.}~\bibnamefont {Liu}}, \bibinfo {author} {\bibfnamefont {T.}~\bibnamefont {Li}}, \bibinfo {author} {\bibfnamefont {S.}~\bibnamefont {Ghosh}},\ and\ \bibinfo {author} {\bibfnamefont {Q.}~\bibnamefont {Xiong}},\ }\bibfield  {title} {\bibinfo {title} {Unraveling the ultrafast coherent dynamics of exciton polariton propagation at room temperature},\ }\href {https://doi.org/10.1021/acs.nanolett.3c02547} {\bibfield  {journal} {\bibinfo  {journal} {Nano Letters}\ }\textbf {\bibinfo {volume} {23}},\ \bibinfo {pages} {8704} (\bibinfo {year} {2023})},\ \bibinfo {note} {pMID: 37681647},\ \Eprint {https://arxiv.org/abs/https://doi.org/10.1021/acs.nanolett.3c02547} {https://doi.org/10.1021/acs.nanolett.3c02547} \BibitemShut {NoStop}%
\bibitem [{\citenamefont {Peng}\ \emph {et~al.}(2022{\natexlab{b}})\citenamefont {Peng}, \citenamefont {Tao}, \citenamefont {Haeberl{\'e}}, \citenamefont {Li}, \citenamefont {Jin}, \citenamefont {Fleming}, \citenamefont {K{\'e}na-Cohen}, \citenamefont {Zhang},\ and\ \citenamefont {Bao}}]{art:peng_RoomT_per}%
  \BibitemOpen
  \bibfield  {author} {\bibinfo {author} {\bibfnamefont {K.}~\bibnamefont {Peng}}, \bibinfo {author} {\bibfnamefont {R.}~\bibnamefont {Tao}}, \bibinfo {author} {\bibfnamefont {L.}~\bibnamefont {Haeberl{\'e}}}, \bibinfo {author} {\bibfnamefont {Q.}~\bibnamefont {Li}}, \bibinfo {author} {\bibfnamefont {D.}~\bibnamefont {Jin}}, \bibinfo {author} {\bibfnamefont {G.~R.}\ \bibnamefont {Fleming}}, \bibinfo {author} {\bibfnamefont {S.}~\bibnamefont {K{\'e}na-Cohen}}, \bibinfo {author} {\bibfnamefont {X.}~\bibnamefont {Zhang}},\ and\ \bibinfo {author} {\bibfnamefont {W.}~\bibnamefont {Bao}},\ }\bibfield  {title} {\bibinfo {title} {Room-temperature polariton quantum fluids in halide perovskites},\ }\href@noop {} {\bibfield  {journal} {\bibinfo  {journal} {Nat. Commun.}\ }\textbf {\bibinfo {volume} {13}},\ \bibinfo {pages} {7388} (\bibinfo {year} {2022}{\natexlab{b}})}\BibitemShut {NoStop}%
\bibitem [{\citenamefont {Wang}\ \emph {et~al.}(2024)\citenamefont {Wang}, \citenamefont {Adamo}, \citenamefont {Ha}, \citenamefont {Tian},\ and\ \citenamefont {Soci}}]{art:Soci_pero}%
  \BibitemOpen
  \bibfield  {author} {\bibinfo {author} {\bibfnamefont {Y.}~\bibnamefont {Wang}}, \bibinfo {author} {\bibfnamefont {G.}~\bibnamefont {Adamo}}, \bibinfo {author} {\bibfnamefont {S.~T.}\ \bibnamefont {Ha}}, \bibinfo {author} {\bibfnamefont {J.}~\bibnamefont {Tian}},\ and\ \bibinfo {author} {\bibfnamefont {C.}~\bibnamefont {Soci}},\ }\href {https://arxiv.org/abs/2404.13256} {\bibinfo {title} {Electrically generated exciton polaritons with spin on-demand}} (\bibinfo {year} {2024}),\ \Eprint {https://arxiv.org/abs/2404.13256} {arXiv:2404.13256 [physics.optics]} \BibitemShut {NoStop}%
\bibitem [{\citenamefont {Neutzner}\ \emph {et~al.}(2018)\citenamefont {Neutzner}, \citenamefont {Thouin}, \citenamefont {Cortecchia}, \citenamefont {Petrozza}, \citenamefont {Silva},\ and\ \citenamefont {Srimath~Kandada}}]{art:neutzner}%
  \BibitemOpen
  \bibfield  {author} {\bibinfo {author} {\bibfnamefont {S.}~\bibnamefont {Neutzner}}, \bibinfo {author} {\bibfnamefont {F.}~\bibnamefont {Thouin}}, \bibinfo {author} {\bibfnamefont {D.}~\bibnamefont {Cortecchia}}, \bibinfo {author} {\bibfnamefont {A.}~\bibnamefont {Petrozza}}, \bibinfo {author} {\bibfnamefont {C.}~\bibnamefont {Silva}},\ and\ \bibinfo {author} {\bibfnamefont {A.~R.}\ \bibnamefont {Srimath~Kandada}},\ }\bibfield  {title} {\bibinfo {title} {Exciton-polaron spectral structures in two-dimensional hybrid lead-halide perovskites},\ }\href {https://doi.org/10.1103/PhysRevMaterials.2.064605} {\bibfield  {journal} {\bibinfo  {journal} {Phys. Rev. Mater.}\ }\textbf {\bibinfo {volume} {2}},\ \bibinfo {pages} {064605} (\bibinfo {year} {2018})}\BibitemShut {NoStop}%
\bibitem [{\citenamefont {Srimath~Kandada}\ \emph {et~al.}(2022)\citenamefont {Srimath~Kandada}, \citenamefont {Li}, \citenamefont {Bittner},\ and\ \citenamefont {Silva-Acuña}}]{Kandada2022}%
  \BibitemOpen
  \bibfield  {author} {\bibinfo {author} {\bibfnamefont {A.~R.}\ \bibnamefont {Srimath~Kandada}}, \bibinfo {author} {\bibfnamefont {H.}~\bibnamefont {Li}}, \bibinfo {author} {\bibfnamefont {E.~R.}\ \bibnamefont {Bittner}},\ and\ \bibinfo {author} {\bibfnamefont {C.}~\bibnamefont {Silva-Acuña}},\ }\bibfield  {title} {\bibinfo {title} {Homogeneous optical line widths in hybrid ruddlesden–popper metal halides can only be measured using nonlinear spectroscopy},\ }\href {https://doi.org/10.1021/acs.jpcc.2c00658} {\bibfield  {journal} {\bibinfo  {journal} {The Journal of Physical Chemistry C}\ }\textbf {\bibinfo {volume} {126}},\ \bibinfo {pages} {5378} (\bibinfo {year} {2022})},\ \Eprint {https://arxiv.org/abs/https://doi.org/10.1021/acs.jpcc.2c00658} {https://doi.org/10.1021/acs.jpcc.2c00658} \BibitemShut {NoStop}%
\bibitem [{\citenamefont {Mermet-Lyaudoz}\ \emph {et~al.}(2023)\citenamefont {Mermet-Lyaudoz}, \citenamefont {Symonds}, \citenamefont {Berry}, \citenamefont {Drouard}, \citenamefont {Chevalier}, \citenamefont {Trippé-Allard}, \citenamefont {Deleporte}, \citenamefont {Bellessa}, \citenamefont {Seassal},\ and\ \citenamefont {Nguyen}}]{Mermet2023}%
  \BibitemOpen
  \bibfield  {author} {\bibinfo {author} {\bibfnamefont {R.}~\bibnamefont {Mermet-Lyaudoz}}, \bibinfo {author} {\bibfnamefont {C.}~\bibnamefont {Symonds}}, \bibinfo {author} {\bibfnamefont {F.}~\bibnamefont {Berry}}, \bibinfo {author} {\bibfnamefont {E.}~\bibnamefont {Drouard}}, \bibinfo {author} {\bibfnamefont {C.}~\bibnamefont {Chevalier}}, \bibinfo {author} {\bibfnamefont {G.}~\bibnamefont {Trippé-Allard}}, \bibinfo {author} {\bibfnamefont {E.}~\bibnamefont {Deleporte}}, \bibinfo {author} {\bibfnamefont {J.}~\bibnamefont {Bellessa}}, \bibinfo {author} {\bibfnamefont {C.}~\bibnamefont {Seassal}},\ and\ \bibinfo {author} {\bibfnamefont {H.~S.}\ \bibnamefont {Nguyen}},\ }\bibfield  {title} {\bibinfo {title} {Taming friedrich–wintgen interference in a resonant metasurface: Vortex laser emitting at an on-demand tilted angle},\ }\href {https://doi.org/10.1021/acs.nanolett.2c04936} {\bibfield  {journal} {\bibinfo  {journal} {Nano Letters}\ }\textbf {\bibinfo {volume} {23}},\ \bibinfo {pages} {4152} (\bibinfo
  {year} {2023})}\BibitemShut {NoStop}%
\bibitem [{\citenamefont {Dang}\ \emph {et~al.}(2024)\citenamefont {Dang}, \citenamefont {Bouteyre}, \citenamefont {Tripp\'{e}-Allard}, \citenamefont {Chevalier}, \citenamefont {Deleporte}, \citenamefont {Drouard}, \citenamefont {Seassal},\ and\ \citenamefont {Nguyen}}]{art:dang2024}%
  \BibitemOpen
  \bibfield  {author} {\bibinfo {author} {\bibfnamefont {N.~H.~M.}\ \bibnamefont {Dang}}, \bibinfo {author} {\bibfnamefont {P.}~\bibnamefont {Bouteyre}}, \bibinfo {author} {\bibfnamefont {G.}~\bibnamefont {Tripp\'{e}-Allard}}, \bibinfo {author} {\bibfnamefont {C.}~\bibnamefont {Chevalier}}, \bibinfo {author} {\bibfnamefont {E.}~\bibnamefont {Deleporte}}, \bibinfo {author} {\bibfnamefont {E.}~\bibnamefont {Drouard}}, \bibinfo {author} {\bibfnamefont {C.}~\bibnamefont {Seassal}},\ and\ \bibinfo {author} {\bibfnamefont {H.~S.}\ \bibnamefont {Nguyen}},\ }\bibfield  {title} {\bibinfo {title} {Nanoimprinted exciton-polaritons metasurfaces: cost-effective, large-scale, high homogeneity, and room temperature operation},\ }\href {https://doi.org/10.1364/OME.512255} {\bibfield  {journal} {\bibinfo  {journal} {Opt. Mater. Express}\ }\textbf {\bibinfo {volume} {14}},\ \bibinfo {pages} {1655} (\bibinfo {year} {2024})}\BibitemShut {NoStop}%
\bibitem [{\citenamefont {Zanotti}\ \emph {et~al.}(2024)\citenamefont {Zanotti}, \citenamefont {Minkov}, \citenamefont {Nigro}, \citenamefont {Gerace}, \citenamefont {Fan},\ and\ \citenamefont {Andreani}}]{art:Zanotti2024}%
  \BibitemOpen
  \bibfield  {author} {\bibinfo {author} {\bibfnamefont {S.}~\bibnamefont {Zanotti}}, \bibinfo {author} {\bibfnamefont {M.}~\bibnamefont {Minkov}}, \bibinfo {author} {\bibfnamefont {D.}~\bibnamefont {Nigro}}, \bibinfo {author} {\bibfnamefont {D.}~\bibnamefont {Gerace}}, \bibinfo {author} {\bibfnamefont {S.}~\bibnamefont {Fan}},\ and\ \bibinfo {author} {\bibfnamefont {L.~C.}\ \bibnamefont {Andreani}},\ }\bibfield  {title} {\bibinfo {title} {Legume: A free implementation of the guided-mode expansion method for photonic crystal slabs},\ }\href {https://doi.org/https://doi.org/10.1016/j.cpc.2024.109286} {\bibfield  {journal} {\bibinfo  {journal} {Computer Physics Communications}\ }\textbf {\bibinfo {volume} {304}},\ \bibinfo {pages} {109286} (\bibinfo {year} {2024})}\BibitemShut {NoStop}%
\bibitem [{rep()}]{repo:legume}%
  \BibitemOpen
  \href@noop {} {\bibinfo {title} {legume: Differentiable guided mode expansion methods}},\ \bibinfo {howpublished} {\url{https://github.com/fancompute/legume}},\ \bibinfo {note} {accessed: 2024-07-15}\BibitemShut {NoStop}%
\bibitem [{\citenamefont {Dang}\ \emph {et~al.}(2020)\citenamefont {Dang}, \citenamefont {Gerace}, \citenamefont {Drouard}, \citenamefont {Tripp{\'{e}}-Allard}, \citenamefont {L{\'{e}}d{\'{e}}e}, \citenamefont {Mazurczyk}, \citenamefont {Deleporte}, \citenamefont {Seassal},\ and\ \citenamefont {Nguyen}}]{Dang2020}%
  \BibitemOpen
  \bibfield  {author} {\bibinfo {author} {\bibfnamefont {N.~H.~M.}\ \bibnamefont {Dang}}, \bibinfo {author} {\bibfnamefont {D.}~\bibnamefont {Gerace}}, \bibinfo {author} {\bibfnamefont {E.}~\bibnamefont {Drouard}}, \bibinfo {author} {\bibfnamefont {G.}~\bibnamefont {Tripp{\'{e}}-Allard}}, \bibinfo {author} {\bibfnamefont {F.}~\bibnamefont {L{\'{e}}d{\'{e}}e}}, \bibinfo {author} {\bibfnamefont {R.}~\bibnamefont {Mazurczyk}}, \bibinfo {author} {\bibfnamefont {E.}~\bibnamefont {Deleporte}}, \bibinfo {author} {\bibfnamefont {C.}~\bibnamefont {Seassal}},\ and\ \bibinfo {author} {\bibfnamefont {H.~S.}\ \bibnamefont {Nguyen}},\ }\bibfield  {title} {\bibinfo {title} {{Tailoring Dispersion of Room-Temperature Exciton-Polaritons with Perovskite-Based Subwavelength Metasurfaces}},\ }\href {https://doi.org/10.1021/acs.nanolett.0c00125} {\bibfield  {journal} {\bibinfo  {journal} {Nano Letters}\ }\textbf {\bibinfo {volume} {20}},\ \bibinfo {pages} {2113} (\bibinfo {year} {2020})}\BibitemShut {NoStop}%
\bibitem [{\citenamefont {Dang}\ \emph {et~al.}(2022)\citenamefont {Dang}, \citenamefont {Zanotti}, \citenamefont {Drouard}, \citenamefont {Chevalier}, \citenamefont {Trippé-Allard}, \citenamefont {Amara}, \citenamefont {Deleporte}, \citenamefont {Ardizzone}, \citenamefont {Sanvitto}, \citenamefont {Andreani}, \citenamefont {Seassal}, \citenamefont {Gerace},\ and\ \citenamefont {Nguyen}}]{art:HaMy2022AOM}%
  \BibitemOpen
  \bibfield  {author} {\bibinfo {author} {\bibfnamefont {N.~H.~M.}\ \bibnamefont {Dang}}, \bibinfo {author} {\bibfnamefont {S.}~\bibnamefont {Zanotti}}, \bibinfo {author} {\bibfnamefont {E.}~\bibnamefont {Drouard}}, \bibinfo {author} {\bibfnamefont {C.}~\bibnamefont {Chevalier}}, \bibinfo {author} {\bibfnamefont {G.}~\bibnamefont {Trippé-Allard}}, \bibinfo {author} {\bibfnamefont {M.}~\bibnamefont {Amara}}, \bibinfo {author} {\bibfnamefont {E.}~\bibnamefont {Deleporte}}, \bibinfo {author} {\bibfnamefont {V.}~\bibnamefont {Ardizzone}}, \bibinfo {author} {\bibfnamefont {D.}~\bibnamefont {Sanvitto}}, \bibinfo {author} {\bibfnamefont {L.~C.}\ \bibnamefont {Andreani}}, \bibinfo {author} {\bibfnamefont {C.}~\bibnamefont {Seassal}}, \bibinfo {author} {\bibfnamefont {D.}~\bibnamefont {Gerace}},\ and\ \bibinfo {author} {\bibfnamefont {H.~S.}\ \bibnamefont {Nguyen}},\ }\bibfield  {title} {\bibinfo {title} {Realization of polaritonic topological charge at room temperature using polariton bound states in the continuum
  from perovskite metasurface},\ }\href {https://doi.org/https://doi.org/10.1002/adom.202102386} {\bibfield  {journal} {\bibinfo  {journal} {Advanced Optical Materials}\ }\textbf {\bibinfo {volume} {10}},\ \bibinfo {pages} {2102386} (\bibinfo {year} {2022})},\ \Eprint {https://arxiv.org/abs/https://onlinelibrary.wiley.com/doi/pdf/10.1002/adom.202102386} {https://onlinelibrary.wiley.com/doi/pdf/10.1002/adom.202102386} \BibitemShut {NoStop}%
\bibitem [{\citenamefont {Savona}\ and\ \citenamefont {Piermarocchi}(1997)}]{art:Savona-Piermarocchi1997}%
  \BibitemOpen
  \bibfield  {author} {\bibinfo {author} {\bibfnamefont {V.}~\bibnamefont {Savona}}\ and\ \bibinfo {author} {\bibfnamefont {C.}~\bibnamefont {Piermarocchi}},\ }\bibfield  {title} {\bibinfo {title} {Microcavity polaritons: Homogeneous and inhomogeneous broadening in the strong coupling regime},\ }\href {https://doi.org/https://doi.org/10.1002/1521-396X(199711)164:1<45::AID-PSSA45>3.0.CO;2-7} {\bibfield  {journal} {\bibinfo  {journal} {physica status solidi (a)}\ }\textbf {\bibinfo {volume} {164}},\ \bibinfo {pages} {45} (\bibinfo {year} {1997})},\ \Eprint {https://arxiv.org/abs/https://onlinelibrary.wiley.com/doi/pdf/10.1002/1521-396X} {https://onlinelibrary.wiley.com/doi/pdf/10.1002/1521-396X} \BibitemShut {NoStop}%
\bibitem [{\citenamefont {Trichet}\ \emph {et~al.}(2011)\citenamefont {Trichet}, \citenamefont {Sun}, \citenamefont {Pavlovic}, \citenamefont {Gippius}, \citenamefont {Malpuech}, \citenamefont {Xie}, \citenamefont {Chen}, \citenamefont {Richard},\ and\ \citenamefont {Dang}}]{art:trichet}%
  \BibitemOpen
  \bibfield  {author} {\bibinfo {author} {\bibfnamefont {A.}~\bibnamefont {Trichet}}, \bibinfo {author} {\bibfnamefont {L.}~\bibnamefont {Sun}}, \bibinfo {author} {\bibfnamefont {G.}~\bibnamefont {Pavlovic}}, \bibinfo {author} {\bibfnamefont {N.}~\bibnamefont {Gippius}}, \bibinfo {author} {\bibfnamefont {G.}~\bibnamefont {Malpuech}}, \bibinfo {author} {\bibfnamefont {W.}~\bibnamefont {Xie}}, \bibinfo {author} {\bibfnamefont {Z.}~\bibnamefont {Chen}}, \bibinfo {author} {\bibfnamefont {M.}~\bibnamefont {Richard}},\ and\ \bibinfo {author} {\bibfnamefont {L.~S.}\ \bibnamefont {Dang}},\ }\bibfield  {title} {\bibinfo {title} {One-dimensional zno exciton polaritons with negligible thermal broadening at room temperature},\ }\href {https://doi.org/10.1103/PhysRevB.83.041302} {\bibfield  {journal} {\bibinfo  {journal} {Phys. Rev. B}\ }\textbf {\bibinfo {volume} {83}},\ \bibinfo {pages} {041302} (\bibinfo {year} {2011})}\BibitemShut {NoStop}%
\bibitem [{\citenamefont {Trichet}(2012)}]{trichetPhD}%
  \BibitemOpen
  \bibfield  {author} {\bibinfo {author} {\bibfnamefont {A.}~\bibnamefont {Trichet}},\ }\emph {\bibinfo {title} {{Polaritons unidimensionnels dans les microfils de Zno : vers la d{\'e}g{\'e}n{\'e}rescence quantique dans les gaz de polaritons unidimensionnels}}},\ \href {https://theses.hal.science/tel-00720629} {\bibinfo {type} {Theses}},\ \bibinfo  {school} {{Universit{\'e} de Grenoble}} (\bibinfo {year} {2012})\BibitemShut {NoStop}%
\bibitem [{\citenamefont {Ferreira}\ \emph {et~al.}(2022)\citenamefont {Ferreira}, \citenamefont {Rosati},\ and\ \citenamefont {Malic}}]{Ferreira2022}%
  \BibitemOpen
  \bibfield  {author} {\bibinfo {author} {\bibfnamefont {B.}~\bibnamefont {Ferreira}}, \bibinfo {author} {\bibfnamefont {R.}~\bibnamefont {Rosati}},\ and\ \bibinfo {author} {\bibfnamefont {E.}~\bibnamefont {Malic}},\ }\bibfield  {title} {\bibinfo {title} {Microscopic modeling of exciton-polariton diffusion coefficients in atomically thin semiconductors},\ }\href {https://doi.org/10.1103/PhysRevMaterials.6.034008} {\bibfield  {journal} {\bibinfo  {journal} {Phys. Rev. Mater.}\ }\textbf {\bibinfo {volume} {6}},\ \bibinfo {pages} {034008} (\bibinfo {year} {2022})}\BibitemShut {NoStop}%
\bibitem [{\citenamefont {Gauthron}\ \emph {et~al.}(2010)\citenamefont {Gauthron}, \citenamefont {Lauret}, \citenamefont {Doyennette}, \citenamefont {Lanty}, \citenamefont {Choueiry}, \citenamefont {Zhang}, \citenamefont {Brehier}, \citenamefont {Largeau}, \citenamefont {Mauguin}, \citenamefont {Bloch},\ and\ \citenamefont {Deleporte}}]{art:gauthron}%
  \BibitemOpen
  \bibfield  {author} {\bibinfo {author} {\bibfnamefont {K.}~\bibnamefont {Gauthron}}, \bibinfo {author} {\bibfnamefont {J.-S.}\ \bibnamefont {Lauret}}, \bibinfo {author} {\bibfnamefont {L.}~\bibnamefont {Doyennette}}, \bibinfo {author} {\bibfnamefont {G.}~\bibnamefont {Lanty}}, \bibinfo {author} {\bibfnamefont {A.~A.}\ \bibnamefont {Choueiry}}, \bibinfo {author} {\bibfnamefont {S.~J.}\ \bibnamefont {Zhang}}, \bibinfo {author} {\bibfnamefont {A.}~\bibnamefont {Brehier}}, \bibinfo {author} {\bibfnamefont {L.}~\bibnamefont {Largeau}}, \bibinfo {author} {\bibfnamefont {O.}~\bibnamefont {Mauguin}}, \bibinfo {author} {\bibfnamefont {J.}~\bibnamefont {Bloch}},\ and\ \bibinfo {author} {\bibfnamefont {E.}~\bibnamefont {Deleporte}},\ }\bibfield  {title} {\bibinfo {title} {Optical spectroscopy of two-dimensional layered (c{\_}6h{\_}5c{\_}2h{\_}4-{NH}{\_}3){\_}2-{PbI}{\_}4 perovskite},\ }\href {https://doi.org/10.1364/oe.18.005912} {\bibfield  {journal} {\bibinfo  {journal} {Optics Express}\ }\textbf {\bibinfo {volume}
  {18}},\ \bibinfo {pages} {5912} (\bibinfo {year} {2010})}\BibitemShut {NoStop}%
\bibitem [{\citenamefont {Feldstein}\ \emph {et~al.}(2020)\citenamefont {Feldstein}, \citenamefont {Perea-Causín}, \citenamefont {Wang}, \citenamefont {Dyksik}, \citenamefont {Watanabe}, \citenamefont {Taniguchi}, \citenamefont {Plochocka},\ and\ \citenamefont {Malic}}]{Feldstein2020}%
  \BibitemOpen
  \bibfield  {author} {\bibinfo {author} {\bibfnamefont {D.}~\bibnamefont {Feldstein}}, \bibinfo {author} {\bibfnamefont {R.}~\bibnamefont {Perea-Causín}}, \bibinfo {author} {\bibfnamefont {S.}~\bibnamefont {Wang}}, \bibinfo {author} {\bibfnamefont {M.}~\bibnamefont {Dyksik}}, \bibinfo {author} {\bibfnamefont {K.}~\bibnamefont {Watanabe}}, \bibinfo {author} {\bibfnamefont {T.}~\bibnamefont {Taniguchi}}, \bibinfo {author} {\bibfnamefont {P.}~\bibnamefont {Plochocka}},\ and\ \bibinfo {author} {\bibfnamefont {E.}~\bibnamefont {Malic}},\ }\bibfield  {title} {\bibinfo {title} {Microscopic picture of electron–phonon interaction in two-dimensional halide perovskites},\ }\href {https://doi.org/10.1021/acs.jpclett.0c02661} {\bibfield  {journal} {\bibinfo  {journal} {The Journal of Physical Chemistry Letters}\ }\textbf {\bibinfo {volume} {11}},\ \bibinfo {pages} {9975} (\bibinfo {year} {2020})},\ \bibinfo {note} {pMID: 33180499},\ \Eprint {https://arxiv.org/abs/https://doi.org/10.1021/acs.jpclett.0c02661}
  {https://doi.org/10.1021/acs.jpclett.0c02661} \BibitemShut {NoStop}%
\bibitem [{\citenamefont {Bao}\ \emph {et~al.}(2019)\citenamefont {Bao}, \citenamefont {Liu}, \citenamefont {Xue}, \citenamefont {Zheng}, \citenamefont {Tao}, \citenamefont {Wang}, \citenamefont {Xia}, \citenamefont {Zhao}, \citenamefont {Kim}, \citenamefont {Yang}, \citenamefont {Li}, \citenamefont {Wang}, \citenamefont {Wang}, \citenamefont {Wang}, \citenamefont {MacDonald},\ and\ \citenamefont {Zhang}}]{Bao2019}%
  \BibitemOpen
  \bibfield  {author} {\bibinfo {author} {\bibfnamefont {W.}~\bibnamefont {Bao}}, \bibinfo {author} {\bibfnamefont {X.}~\bibnamefont {Liu}}, \bibinfo {author} {\bibfnamefont {F.}~\bibnamefont {Xue}}, \bibinfo {author} {\bibfnamefont {F.}~\bibnamefont {Zheng}}, \bibinfo {author} {\bibfnamefont {R.}~\bibnamefont {Tao}}, \bibinfo {author} {\bibfnamefont {S.}~\bibnamefont {Wang}}, \bibinfo {author} {\bibfnamefont {Y.}~\bibnamefont {Xia}}, \bibinfo {author} {\bibfnamefont {M.}~\bibnamefont {Zhao}}, \bibinfo {author} {\bibfnamefont {J.}~\bibnamefont {Kim}}, \bibinfo {author} {\bibfnamefont {S.}~\bibnamefont {Yang}}, \bibinfo {author} {\bibfnamefont {Q.}~\bibnamefont {Li}}, \bibinfo {author} {\bibfnamefont {Y.}~\bibnamefont {Wang}}, \bibinfo {author} {\bibfnamefont {Y.}~\bibnamefont {Wang}}, \bibinfo {author} {\bibfnamefont {L.-W.}\ \bibnamefont {Wang}}, \bibinfo {author} {\bibfnamefont {A.~H.}\ \bibnamefont {MacDonald}},\ and\ \bibinfo {author} {\bibfnamefont {X.}~\bibnamefont {Zhang}},\ }\bibfield  {title}
  {\bibinfo {title} {Observation of rydberg exciton polaritons and their condensate in a perovskite cavity},\ }\href {https://doi.org/10.1073/pnas.1909948116} {\bibfield  {journal} {\bibinfo  {journal} {Proceedings of the National Academy of Sciences}\ }\textbf {\bibinfo {volume} {116}},\ \bibinfo {pages} {20274–20279} (\bibinfo {year} {2019})}\BibitemShut {NoStop}%
\bibitem [{\citenamefont {Wang}\ \emph {et~al.}(2018)\citenamefont {Wang}, \citenamefont {Su}, \citenamefont {Xing}, \citenamefont {Bao}, \citenamefont {Diederichs}, \citenamefont {Liu}, \citenamefont {Liew}, \citenamefont {Chen},\ and\ \citenamefont {Xiong}}]{Wang2018}%
  \BibitemOpen
  \bibfield  {author} {\bibinfo {author} {\bibfnamefont {J.}~\bibnamefont {Wang}}, \bibinfo {author} {\bibfnamefont {R.}~\bibnamefont {Su}}, \bibinfo {author} {\bibfnamefont {J.}~\bibnamefont {Xing}}, \bibinfo {author} {\bibfnamefont {D.}~\bibnamefont {Bao}}, \bibinfo {author} {\bibfnamefont {C.}~\bibnamefont {Diederichs}}, \bibinfo {author} {\bibfnamefont {S.}~\bibnamefont {Liu}}, \bibinfo {author} {\bibfnamefont {T.~C.}\ \bibnamefont {Liew}}, \bibinfo {author} {\bibfnamefont {Z.}~\bibnamefont {Chen}},\ and\ \bibinfo {author} {\bibfnamefont {Q.}~\bibnamefont {Xiong}},\ }\bibfield  {title} {\bibinfo {title} {Room temperature coherently coupled exciton–polaritons in two-dimensional organic–inorganic perovskite},\ }\href {https://doi.org/10.1021/acsnano.8b03737} {\bibfield  {journal} {\bibinfo  {journal} {ACS Nano}\ }\textbf {\bibinfo {volume} {12}},\ \bibinfo {pages} {8382–8389} (\bibinfo {year} {2018})}\BibitemShut {NoStop}%
\bibitem [{\citenamefont {Whittaker}\ \emph {et~al.}(1996)\citenamefont {Whittaker}, \citenamefont {Kinsler}, \citenamefont {Fisher}, \citenamefont {Skolnick}, \citenamefont {Armitage}, \citenamefont {Afshar}, \citenamefont {Sturge},\ and\ \citenamefont {Roberts}}]{Whittaker1996}%
  \BibitemOpen
  \bibfield  {author} {\bibinfo {author} {\bibfnamefont {D.~M.}\ \bibnamefont {Whittaker}}, \bibinfo {author} {\bibfnamefont {P.}~\bibnamefont {Kinsler}}, \bibinfo {author} {\bibfnamefont {T.~A.}\ \bibnamefont {Fisher}}, \bibinfo {author} {\bibfnamefont {M.~S.}\ \bibnamefont {Skolnick}}, \bibinfo {author} {\bibfnamefont {A.}~\bibnamefont {Armitage}}, \bibinfo {author} {\bibfnamefont {A.~M.}\ \bibnamefont {Afshar}}, \bibinfo {author} {\bibfnamefont {M.~D.}\ \bibnamefont {Sturge}},\ and\ \bibinfo {author} {\bibfnamefont {J.~S.}\ \bibnamefont {Roberts}},\ }\bibfield  {title} {\bibinfo {title} {Motional narrowing in semiconductor microcavities},\ }\href {https://doi.org/10.1103/physrevlett.77.4792} {\bibfield  {journal} {\bibinfo  {journal} {Physical Review Letters}\ }\textbf {\bibinfo {volume} {77}},\ \bibinfo {pages} {4792–4795} (\bibinfo {year} {1996})}\BibitemShut {NoStop}%
\bibitem [{\citenamefont {Savona}\ \emph {et~al.}(1997)\citenamefont {Savona}, \citenamefont {Piermarocchi}, \citenamefont {Quattropani}, \citenamefont {Tassone},\ and\ \citenamefont {Schwendimann}}]{Savona1997}%
  \BibitemOpen
  \bibfield  {author} {\bibinfo {author} {\bibfnamefont {V.}~\bibnamefont {Savona}}, \bibinfo {author} {\bibfnamefont {C.}~\bibnamefont {Piermarocchi}}, \bibinfo {author} {\bibfnamefont {A.}~\bibnamefont {Quattropani}}, \bibinfo {author} {\bibfnamefont {F.}~\bibnamefont {Tassone}},\ and\ \bibinfo {author} {\bibfnamefont {P.}~\bibnamefont {Schwendimann}},\ }\bibfield  {title} {\bibinfo {title} {Microscopic theory of motional narrowing of microcavity polaritons in a disordered potential},\ }\href {https://doi.org/10.1103/physrevlett.78.4470} {\bibfield  {journal} {\bibinfo  {journal} {Physical Review Letters}\ }\textbf {\bibinfo {volume} {78}},\ \bibinfo {pages} {4470–4473} (\bibinfo {year} {1997})}\BibitemShut {NoStop}%
\bibitem [{\citenamefont {Kravtsov}\ \emph {et~al.}(2020)\citenamefont {Kravtsov}, \citenamefont {Khestanova}, \citenamefont {Benimetskiy}, \citenamefont {Ivanova}, \citenamefont {Samusev}, \citenamefont {Sinev}, \citenamefont {Pidgayko}, \citenamefont {Mozharov}, \citenamefont {Mukhin}, \citenamefont {Lozhkin}, \citenamefont {Kapitonov}, \citenamefont {Brichkin}, \citenamefont {Kulakovskii}, \citenamefont {Shelykh}, \citenamefont {Tartakovskii}, \citenamefont {Walker}, \citenamefont {Skolnick}, \citenamefont {Krizhanovskii},\ and\ \citenamefont {Iorsh}}]{Kravtsov2020}%
  \BibitemOpen
  \bibfield  {author} {\bibinfo {author} {\bibfnamefont {V.}~\bibnamefont {Kravtsov}}, \bibinfo {author} {\bibfnamefont {E.}~\bibnamefont {Khestanova}}, \bibinfo {author} {\bibfnamefont {F.~A.}\ \bibnamefont {Benimetskiy}}, \bibinfo {author} {\bibfnamefont {T.}~\bibnamefont {Ivanova}}, \bibinfo {author} {\bibfnamefont {A.~K.}\ \bibnamefont {Samusev}}, \bibinfo {author} {\bibfnamefont {I.~S.}\ \bibnamefont {Sinev}}, \bibinfo {author} {\bibfnamefont {D.}~\bibnamefont {Pidgayko}}, \bibinfo {author} {\bibfnamefont {A.~M.}\ \bibnamefont {Mozharov}}, \bibinfo {author} {\bibfnamefont {I.~S.}\ \bibnamefont {Mukhin}}, \bibinfo {author} {\bibfnamefont {M.~S.}\ \bibnamefont {Lozhkin}}, \bibinfo {author} {\bibfnamefont {Y.~V.}\ \bibnamefont {Kapitonov}}, \bibinfo {author} {\bibfnamefont {A.~S.}\ \bibnamefont {Brichkin}}, \bibinfo {author} {\bibfnamefont {V.~D.}\ \bibnamefont {Kulakovskii}}, \bibinfo {author} {\bibfnamefont {I.~A.}\ \bibnamefont {Shelykh}}, \bibinfo {author} {\bibfnamefont {A.~I.}\ \bibnamefont
  {Tartakovskii}}, \bibinfo {author} {\bibfnamefont {P.~M.}\ \bibnamefont {Walker}}, \bibinfo {author} {\bibfnamefont {M.~S.}\ \bibnamefont {Skolnick}}, \bibinfo {author} {\bibfnamefont {D.~N.}\ \bibnamefont {Krizhanovskii}},\ and\ \bibinfo {author} {\bibfnamefont {I.~V.}\ \bibnamefont {Iorsh}},\ }\bibfield  {title} {\bibinfo {title} {Nonlinear polaritons in a monolayer semiconductor coupled to optical bound states in the continuum},\ }\href {https://doi.org/10.1038/s41377-020-0286-z} {\bibfield  {journal} {\bibinfo  {journal} {Light: Science \&; Applications}\ }\textbf {\bibinfo {volume} {9}},\ \bibinfo {pages} {56} (\bibinfo {year} {2020})}\BibitemShut {NoStop}%
\bibitem [{\citenamefont {Wurdack}\ \emph {et~al.}(2021)\citenamefont {Wurdack}, \citenamefont {Estrecho}, \citenamefont {Todd}, \citenamefont {Yun}, \citenamefont {Pieczarka}, \citenamefont {Earl}, \citenamefont {Davis}, \citenamefont {Schneider}, \citenamefont {Truscott},\ and\ \citenamefont {Ostrovskaya}}]{Wurdack2021}%
  \BibitemOpen
  \bibfield  {author} {\bibinfo {author} {\bibfnamefont {M.}~\bibnamefont {Wurdack}}, \bibinfo {author} {\bibfnamefont {E.}~\bibnamefont {Estrecho}}, \bibinfo {author} {\bibfnamefont {S.}~\bibnamefont {Todd}}, \bibinfo {author} {\bibfnamefont {T.}~\bibnamefont {Yun}}, \bibinfo {author} {\bibfnamefont {M.}~\bibnamefont {Pieczarka}}, \bibinfo {author} {\bibfnamefont {S.~K.}\ \bibnamefont {Earl}}, \bibinfo {author} {\bibfnamefont {J.~A.}\ \bibnamefont {Davis}}, \bibinfo {author} {\bibfnamefont {C.}~\bibnamefont {Schneider}}, \bibinfo {author} {\bibfnamefont {A.~G.}\ \bibnamefont {Truscott}},\ and\ \bibinfo {author} {\bibfnamefont {E.~A.}\ \bibnamefont {Ostrovskaya}},\ }\bibfield  {title} {\bibinfo {title} {Motional narrowing, ballistic transport, and trapping of room-temperature exciton polaritons in an atomically-thin semiconductor},\ }\href {https://doi.org/10.1038/s41467-021-25656-7} {\bibfield  {journal} {\bibinfo  {journal} {Nature Communications}\ }\textbf {\bibinfo {volume} {12}},\ \bibinfo {pages} {5366}
  (\bibinfo {year} {2021})}\BibitemShut {NoStop}%
\bibitem [{\citenamefont {Verdelli}\ \emph {et~al.}(2024)\citenamefont {Verdelli}, \citenamefont {Baldi},\ and\ \citenamefont {Rivas}}]{Verdelli2024}%
  \BibitemOpen
  \bibfield  {author} {\bibinfo {author} {\bibfnamefont {F.}~\bibnamefont {Verdelli}}, \bibinfo {author} {\bibfnamefont {A.}~\bibnamefont {Baldi}},\ and\ \bibinfo {author} {\bibfnamefont {J.~G.}\ \bibnamefont {Rivas}},\ }\bibfield  {title} {\bibinfo {title} {Motional narrowing of molecular vibrations strongly coupled to surface lattice resonances},\ }\href {https://doi.org/10.1103/PhysRevB.109.174305} {\bibfield  {journal} {\bibinfo  {journal} {Phys. Rev. B}\ }\textbf {\bibinfo {volume} {109}},\ \bibinfo {pages} {174305} (\bibinfo {year} {2024})}\BibitemShut {NoStop}%
\bibitem [{\citenamefont {Cueff}\ \emph {et~al.}(2024)\citenamefont {Cueff}, \citenamefont {Berguiga},\ and\ \citenamefont {Nguyen}}]{Cueff2024}%
  \BibitemOpen
  \bibfield  {author} {\bibinfo {author} {\bibfnamefont {S.}~\bibnamefont {Cueff}}, \bibinfo {author} {\bibfnamefont {L.}~\bibnamefont {Berguiga}},\ and\ \bibinfo {author} {\bibfnamefont {H.~S.}\ \bibnamefont {Nguyen}},\ }\bibfield  {title} {\bibinfo {title} {Fourier imaging for nanophotonics},\ }\href {https://doi.org/doi:10.1515/nanoph-2023-0887} {\bibfield  {journal} {\bibinfo  {journal} {Nanophotonics}\ }\textbf {\bibinfo {volume} {13}},\ \bibinfo {pages} {841} (\bibinfo {year} {2024})}\BibitemShut {NoStop}%
\bibitem [{\citenamefont {Hou}\ \emph {et~al.}(2020{\natexlab{b}})\citenamefont {Hou}, \citenamefont {Khatoniar}, \citenamefont {Ding}, \citenamefont {Qu}, \citenamefont {Napolov}, \citenamefont {Menon},\ and\ \citenamefont {Forrest}}]{Hou2020}%
  \BibitemOpen
  \bibfield  {author} {\bibinfo {author} {\bibfnamefont {S.}~\bibnamefont {Hou}}, \bibinfo {author} {\bibfnamefont {M.}~\bibnamefont {Khatoniar}}, \bibinfo {author} {\bibfnamefont {K.}~\bibnamefont {Ding}}, \bibinfo {author} {\bibfnamefont {Y.}~\bibnamefont {Qu}}, \bibinfo {author} {\bibfnamefont {A.}~\bibnamefont {Napolov}}, \bibinfo {author} {\bibfnamefont {V.~M.}\ \bibnamefont {Menon}},\ and\ \bibinfo {author} {\bibfnamefont {S.~R.}\ \bibnamefont {Forrest}},\ }\bibfield  {title} {\bibinfo {title} {Ultralong‐range energy transport in a disordered organic semiconductor at room temperature via coherent exciton‐polariton propagation},\ }\href {https://doi.org/10.1002/adma.202002127} {\bibfield  {journal} {\bibinfo  {journal} {Advanced Materials}\ }\textbf {\bibinfo {volume} {32}},\ \bibinfo {pages} {2002127} (\bibinfo {year} {2020}{\natexlab{b}})}\BibitemShut {NoStop}%
\bibitem [{\citenamefont {Ziegler}\ \emph {et~al.}(2020)\citenamefont {Ziegler}, \citenamefont {Zipfel}, \citenamefont {Meisinger}, \citenamefont {Menahem}, \citenamefont {Zhu}, \citenamefont {Taniguchi}, \citenamefont {Watanabe}, \citenamefont {Yaffe}, \citenamefont {Egger},\ and\ \citenamefont {Chernikov}}]{Ziegler2020}%
  \BibitemOpen
  \bibfield  {author} {\bibinfo {author} {\bibfnamefont {J.~D.}\ \bibnamefont {Ziegler}}, \bibinfo {author} {\bibfnamefont {J.}~\bibnamefont {Zipfel}}, \bibinfo {author} {\bibfnamefont {B.}~\bibnamefont {Meisinger}}, \bibinfo {author} {\bibfnamefont {M.}~\bibnamefont {Menahem}}, \bibinfo {author} {\bibfnamefont {X.}~\bibnamefont {Zhu}}, \bibinfo {author} {\bibfnamefont {T.}~\bibnamefont {Taniguchi}}, \bibinfo {author} {\bibfnamefont {K.}~\bibnamefont {Watanabe}}, \bibinfo {author} {\bibfnamefont {O.}~\bibnamefont {Yaffe}}, \bibinfo {author} {\bibfnamefont {D.~A.}\ \bibnamefont {Egger}},\ and\ \bibinfo {author} {\bibfnamefont {A.}~\bibnamefont {Chernikov}},\ }\bibfield  {title} {\bibinfo {title} {Fast and anomalous exciton diffusion in two-dimensional hybrid perovskites},\ }\href {https://doi.org/10.1021/acs.nanolett.0c02472} {\bibfield  {journal} {\bibinfo  {journal} {Nano Letters}\ }\textbf {\bibinfo {volume} {20}},\ \bibinfo {pages} {6674–6681} (\bibinfo {year} {2020})}\BibitemShut {NoStop}%
\bibitem [{\citenamefont {Seitz}\ \emph {et~al.}(2020)\citenamefont {Seitz}, \citenamefont {Magdaleno}, \citenamefont {Alcázar-Cano}, \citenamefont {Meléndez}, \citenamefont {Lubbers}, \citenamefont {Walraven}, \citenamefont {Pakdel}, \citenamefont {Prada}, \citenamefont {Delgado-Buscalioni},\ and\ \citenamefont {Prins}}]{Seitz2020}%
  \BibitemOpen
  \bibfield  {author} {\bibinfo {author} {\bibfnamefont {M.}~\bibnamefont {Seitz}}, \bibinfo {author} {\bibfnamefont {A.~J.}\ \bibnamefont {Magdaleno}}, \bibinfo {author} {\bibfnamefont {N.}~\bibnamefont {Alcázar-Cano}}, \bibinfo {author} {\bibfnamefont {M.}~\bibnamefont {Meléndez}}, \bibinfo {author} {\bibfnamefont {T.~J.}\ \bibnamefont {Lubbers}}, \bibinfo {author} {\bibfnamefont {S.~W.}\ \bibnamefont {Walraven}}, \bibinfo {author} {\bibfnamefont {S.}~\bibnamefont {Pakdel}}, \bibinfo {author} {\bibfnamefont {E.}~\bibnamefont {Prada}}, \bibinfo {author} {\bibfnamefont {R.}~\bibnamefont {Delgado-Buscalioni}},\ and\ \bibinfo {author} {\bibfnamefont {F.}~\bibnamefont {Prins}},\ }\bibfield  {title} {\bibinfo {title} {Exciton diffusion in two-dimensional metal-halide perovskites},\ }\href {https://doi.org/10.1038/s41467-020-15882-w} {\bibfield  {journal} {\bibinfo  {journal} {Nature Communications}\ }\textbf {\bibinfo {volume} {11}},\ \bibinfo {pages} {2035} (\bibinfo {year} {2020})}\BibitemShut {NoStop}%
\bibitem [{\citenamefont {Xiao}\ \emph {et~al.}(2020)\citenamefont {Xiao}, \citenamefont {Wu}, \citenamefont {Ni}, \citenamefont {Xu}, \citenamefont {Chen}, \citenamefont {Hu}, \citenamefont {Rudd}, \citenamefont {You},\ and\ \citenamefont {Huang}}]{Xiao2020}%
  \BibitemOpen
  \bibfield  {author} {\bibinfo {author} {\bibfnamefont {X.}~\bibnamefont {Xiao}}, \bibinfo {author} {\bibfnamefont {M.}~\bibnamefont {Wu}}, \bibinfo {author} {\bibfnamefont {Z.}~\bibnamefont {Ni}}, \bibinfo {author} {\bibfnamefont {S.}~\bibnamefont {Xu}}, \bibinfo {author} {\bibfnamefont {S.}~\bibnamefont {Chen}}, \bibinfo {author} {\bibfnamefont {J.}~\bibnamefont {Hu}}, \bibinfo {author} {\bibfnamefont {P.~N.}\ \bibnamefont {Rudd}}, \bibinfo {author} {\bibfnamefont {W.}~\bibnamefont {You}},\ and\ \bibinfo {author} {\bibfnamefont {J.}~\bibnamefont {Huang}},\ }\bibfield  {title} {\bibinfo {title} {Ultrafast exciton transport with a long diffusion length in layered perovskites with organic cation functionalization},\ }\href {https://doi.org/https://doi.org/10.1002/adma.202004080} {\bibfield  {journal} {\bibinfo  {journal} {Advanced Materials}\ }\textbf {\bibinfo {volume} {32}},\ \bibinfo {pages} {2004080} (\bibinfo {year} {2020})},\ \Eprint
  {https://arxiv.org/abs/https://onlinelibrary.wiley.com/doi/pdf/10.1002/adma.202004080} {https://onlinelibrary.wiley.com/doi/pdf/10.1002/adma.202004080} \BibitemShut {NoStop}%
\bibitem [{\citenamefont {Fieramosca}\ \emph {et~al.}(2019)\citenamefont {Fieramosca}, \citenamefont {Polimeno}, \citenamefont {Ardizzone}, \citenamefont {Marco}, \citenamefont {Pugliese}, \citenamefont {Maiorano}, \citenamefont {Giorgi}, \citenamefont {Dominici}, \citenamefont {Gigli}, \citenamefont {Gerace}, \citenamefont {Ballarini},\ and\ \citenamefont {Sanvitto}}]{Fieramosca2019}%
  \BibitemOpen
  \bibfield  {author} {\bibinfo {author} {\bibfnamefont {A.}~\bibnamefont {Fieramosca}}, \bibinfo {author} {\bibfnamefont {L.}~\bibnamefont {Polimeno}}, \bibinfo {author} {\bibfnamefont {V.}~\bibnamefont {Ardizzone}}, \bibinfo {author} {\bibfnamefont {L.~D.}\ \bibnamefont {Marco}}, \bibinfo {author} {\bibfnamefont {M.}~\bibnamefont {Pugliese}}, \bibinfo {author} {\bibfnamefont {V.}~\bibnamefont {Maiorano}}, \bibinfo {author} {\bibfnamefont {M.~D.}\ \bibnamefont {Giorgi}}, \bibinfo {author} {\bibfnamefont {L.}~\bibnamefont {Dominici}}, \bibinfo {author} {\bibfnamefont {G.}~\bibnamefont {Gigli}}, \bibinfo {author} {\bibfnamefont {D.}~\bibnamefont {Gerace}}, \bibinfo {author} {\bibfnamefont {D.}~\bibnamefont {Ballarini}},\ and\ \bibinfo {author} {\bibfnamefont {D.}~\bibnamefont {Sanvitto}},\ }\bibfield  {title} {\bibinfo {title} {Two-dimensional hybrid perovskites sustaining strong polariton interactions at room temperature},\ }\href {https://doi.org/10.1126/sciadv.aav9967} {\bibfield  {journal} {\bibinfo
  {journal} {Science Advances}\ }\textbf {\bibinfo {volume} {5}},\ \bibinfo {pages} {eaav9967} (\bibinfo {year} {2019})},\ \Eprint {https://arxiv.org/abs/https://www.science.org/doi/pdf/10.1126/sciadv.aav9967} {https://www.science.org/doi/pdf/10.1126/sciadv.aav9967} \BibitemShut {NoStop}%
\bibitem [{\citenamefont {Nguyen}\ \emph {et~al.}(2018)\citenamefont {Nguyen}, \citenamefont {Dubois}, \citenamefont {Deschamps}, \citenamefont {Cueff}, \citenamefont {Pardon}, \citenamefont {Leclercq}, \citenamefont {Seassal}, \citenamefont {Letartre},\ and\ \citenamefont {Viktorovitch}}]{Nguyen2018}%
  \BibitemOpen
  \bibfield  {author} {\bibinfo {author} {\bibfnamefont {H.~S.}\ \bibnamefont {Nguyen}}, \bibinfo {author} {\bibfnamefont {F.}~\bibnamefont {Dubois}}, \bibinfo {author} {\bibfnamefont {T.}~\bibnamefont {Deschamps}}, \bibinfo {author} {\bibfnamefont {S.}~\bibnamefont {Cueff}}, \bibinfo {author} {\bibfnamefont {A.}~\bibnamefont {Pardon}}, \bibinfo {author} {\bibfnamefont {J.-L.}\ \bibnamefont {Leclercq}}, \bibinfo {author} {\bibfnamefont {C.}~\bibnamefont {Seassal}}, \bibinfo {author} {\bibfnamefont {X.}~\bibnamefont {Letartre}},\ and\ \bibinfo {author} {\bibfnamefont {P.}~\bibnamefont {Viktorovitch}},\ }\bibfield  {title} {\bibinfo {title} {Symmetry breaking in photonic crystals: On-demand dispersion from flatband to dirac cones},\ }\href {https://doi.org/10.1103/PhysRevLett.120.066102} {\bibfield  {journal} {\bibinfo  {journal} {Phys. Rev. Lett.}\ }\textbf {\bibinfo {volume} {120}},\ \bibinfo {pages} {066102} (\bibinfo {year} {2018})}\BibitemShut {NoStop}%
\bibitem [{\citenamefont {Zanotti}\ \emph {et~al.}(2022)\citenamefont {Zanotti}, \citenamefont {Nguyen}, \citenamefont {Minkov}, \citenamefont {Andreani},\ and\ \citenamefont {Gerace}}]{art:zanotti}%
  \BibitemOpen
  \bibfield  {author} {\bibinfo {author} {\bibfnamefont {S.}~\bibnamefont {Zanotti}}, \bibinfo {author} {\bibfnamefont {H.~S.}\ \bibnamefont {Nguyen}}, \bibinfo {author} {\bibfnamefont {M.}~\bibnamefont {Minkov}}, \bibinfo {author} {\bibfnamefont {L.~C.}\ \bibnamefont {Andreani}},\ and\ \bibinfo {author} {\bibfnamefont {D.}~\bibnamefont {Gerace}},\ }\bibfield  {title} {\bibinfo {title} {Theory of photonic crystal polaritons in periodically patterned multilayer waveguides},\ }\href {https://doi.org/10.1103/PhysRevB.106.115424} {\bibfield  {journal} {\bibinfo  {journal} {Phys. Rev. B}\ }\textbf {\bibinfo {volume} {106}},\ \bibinfo {pages} {115424} (\bibinfo {year} {2022})}\BibitemShut {NoStop}%
\bibitem [{\citenamefont {Nigro}\ and\ \citenamefont {Gerace}(2023)}]{Nigro2023}%
  \BibitemOpen
  \bibfield  {author} {\bibinfo {author} {\bibfnamefont {D.}~\bibnamefont {Nigro}}\ and\ \bibinfo {author} {\bibfnamefont {D.}~\bibnamefont {Gerace}},\ }\bibfield  {title} {\bibinfo {title} {Theory of exciton-polariton condensation in gap-confined eigenmodes},\ }\href {https://doi.org/10.1103/PhysRevB.108.085305} {\bibfield  {journal} {\bibinfo  {journal} {Phys. Rev. B}\ }\textbf {\bibinfo {volume} {108}},\ \bibinfo {pages} {085305} (\bibinfo {year} {2023})}\BibitemShut {NoStop}%
\bibitem [{\citenamefont {Sigurdsson}\ \emph {et~al.}(2024)\citenamefont {Sigurdsson}, \citenamefont {Nguyen},\ and\ \citenamefont {Nguyen}}]{Sigurdsson2024}%
  \BibitemOpen
  \bibfield  {author} {\bibinfo {author} {\bibfnamefont {H.}~\bibnamefont {Sigurdsson}}, \bibinfo {author} {\bibfnamefont {H.~C.}\ \bibnamefont {Nguyen}},\ and\ \bibinfo {author} {\bibfnamefont {H.~S.}\ \bibnamefont {Nguyen}},\ }\bibfield  {title} {\bibinfo {title} {Dirac exciton–polariton condensates in photonic crystal gratings},\ }\href {https://doi.org/doi:10.1515/nanoph-2023-0834} {\bibfield  {journal} {\bibinfo  {journal} {Nanophotonics}\ ,\ \bibinfo {pages} {https://doi.org/10.1515/nanoph}} (\bibinfo {year} {2024})}\BibitemShut {NoStop}%
\bibitem [{\citenamefont {Wang}\ \emph {et~al.}(2022)\citenamefont {Wang}, \citenamefont {Zang}, \citenamefont {Gao}, \citenamefont {Lyu}, \citenamefont {Gu}, \citenamefont {Yao}, \citenamefont {Peng}, \citenamefont {Watanabe}, \citenamefont {Taniguchi}, \citenamefont {Liu}, \citenamefont {Gao}, \citenamefont {Bao},\ and\ \citenamefont {Ye}}]{art:Elettr_Pump}%
  \BibitemOpen
  \bibfield  {author} {\bibinfo {author} {\bibfnamefont {T.}~\bibnamefont {Wang}}, \bibinfo {author} {\bibfnamefont {Z.}~\bibnamefont {Zang}}, \bibinfo {author} {\bibfnamefont {Y.}~\bibnamefont {Gao}}, \bibinfo {author} {\bibfnamefont {C.}~\bibnamefont {Lyu}}, \bibinfo {author} {\bibfnamefont {P.}~\bibnamefont {Gu}}, \bibinfo {author} {\bibfnamefont {Y.}~\bibnamefont {Yao}}, \bibinfo {author} {\bibfnamefont {K.}~\bibnamefont {Peng}}, \bibinfo {author} {\bibfnamefont {K.}~\bibnamefont {Watanabe}}, \bibinfo {author} {\bibfnamefont {T.}~\bibnamefont {Taniguchi}}, \bibinfo {author} {\bibfnamefont {X.}~\bibnamefont {Liu}}, \bibinfo {author} {\bibfnamefont {Y.}~\bibnamefont {Gao}}, \bibinfo {author} {\bibfnamefont {W.}~\bibnamefont {Bao}},\ and\ \bibinfo {author} {\bibfnamefont {Y.}~\bibnamefont {Ye}},\ }\bibfield  {title} {\bibinfo {title} {Electrically pumped polarized exciton-polaritons in a halide perovskite microcavity},\ }\href {https://doi.org/10.1021/acs.nanolett.2c00906} {\bibfield  {journal} {\bibinfo
  {journal} {Nano Letters}\ }\textbf {\bibinfo {volume} {22}},\ \bibinfo {pages} {5175} (\bibinfo {year} {2022})},\ \Eprint {https://arxiv.org/abs/https://doi.org/10.1021/acs.nanolett.2c00906} {https://doi.org/10.1021/acs.nanolett.2c00906} \BibitemShut {NoStop}%
\bibitem [{\citenamefont {Wang}\ \emph {et~al.}(2023)\citenamefont {Wang}, \citenamefont {Tian}, \citenamefont {Klein}, \citenamefont {Adamo}, \citenamefont {Ha},\ and\ \citenamefont {Soci}}]{art:Wang_El_Pump}%
  \BibitemOpen
  \bibfield  {author} {\bibinfo {author} {\bibfnamefont {Y.}~\bibnamefont {Wang}}, \bibinfo {author} {\bibfnamefont {J.}~\bibnamefont {Tian}}, \bibinfo {author} {\bibfnamefont {M.}~\bibnamefont {Klein}}, \bibinfo {author} {\bibfnamefont {G.}~\bibnamefont {Adamo}}, \bibinfo {author} {\bibfnamefont {S.~T.}\ \bibnamefont {Ha}},\ and\ \bibinfo {author} {\bibfnamefont {C.}~\bibnamefont {Soci}},\ }\bibfield  {title} {\bibinfo {title} {Directional emission from electrically injected exciton–polaritons in perovskite metasurfaces},\ }\href {https://doi.org/10.1021/acs.nanolett.3c00727} {\bibfield  {journal} {\bibinfo  {journal} {Nano Letters}\ }\textbf {\bibinfo {volume} {23}},\ \bibinfo {pages} {4431} (\bibinfo {year} {2023})},\ \bibinfo {note} {pMID: 37129264},\ \Eprint {https://arxiv.org/abs/https://doi.org/10.1021/acs.nanolett.3c00727} {https://doi.org/10.1021/acs.nanolett.3c00727} \BibitemShut {NoStop}%
\end{thebibliography}%
\begin{center}
	\textbf{\large --- SUPPLEMENTAL MATERIAL ---}
\end{center}

\newcommand{\figref}[1]{Fig.~\ref{#1}}
\newcommand{\secref}[1]{Sec.~\ref{#1}}
\renewcommand{\eqref}[1]{Eq.~(\ref{#1})}
\renewcommand{\vec}[1]{\mathbf{#1}}
\newcommand{\vecg}[1]{\boldsymbol{#1}}
\newcommand{\mat}[1]{\mathbb{#1}}
\renewcommand{\appendixname}{APPENDIX}
\newcommand{\appendixsec}[2]{\section{\uppercase{#1}}\label{#2}}
\newcommand{\appendixsubsec}[2]{\subsection{\uppercase{#1}}\label{#2}}
\renewcommand{\mat}[1]{\overline{\overline{#1}}}

\renewcommand{\theequation}{S\arabic{equation}}
\renewcommand{\thefigure}{S\arabic{figure}}

\setcounter{page}{1}
\setcounter{equation}{0}
\setcounter{figure}{0}
\setcounter{table}{0}
\setcounter{section}{0}

\section{Optical Properties of PEPI thin film}

\begin{figure}[h!]
\centering
    \includegraphics[width=\linewidth]{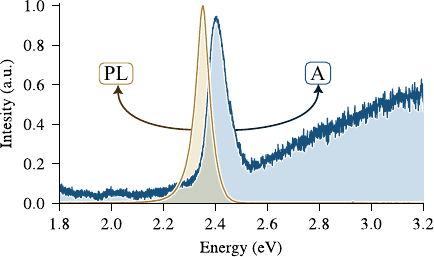}
    \caption{Normalized photoluminescence and absorption spectra of PEPI thin-film. }
    \label{fig:PL_vs_Abs}
\end{figure}
{
In this section, we present the photoluminescence (PL) and absorption (A) spectra measured from a thin PEPI film. The film was obtained by spin-coating the perovskite onto a glass substrate, and its thickness is approximately $D_{\text{film}}=\SI{50}{\nano\meter}$. From the A spectrum reported in Fig.~\ref{fig:PL_vs_Abs}, we can extract the bare exciton energy $E_X=\SI{2.394}{\electronvolt}$. From the red shift between the PL peak with respect to the the absorption peak we estimate a Stokes shift of $\SI{42}{\milli\electronvolt}$.
}

\section{Suppression of thermal broadening  }
In order to investigate the impact of exciton losses on the polariton linewidth, we performed the fit of Eq.~(1) {reported in the main text} in two different configurations. In the first case, we fixed the excitonic linewidth as $\Gamma_X=\SI{0.3}{\milli\electronvolt}$, representing excitonic losses without any thermal broadening effect. Secondly, we set the exciton linewidth corresponding to the bare exciton as $\Gamma_X=\SI{30}{\milli\electronvolt}$, i.e., including thermal broadening effects. The polaritonic linewidths resulting from the two fits are compared in Fig.~\ref{fig:loss_suppr}, in which experimental data extracted from the PL spectrum of Fig.~1(d) in the main text are also included. Remarkably, we observed a significant agreement between the fitted model and experimental results for the choice $\Gamma_X=\SI{0.3}{\milli\electronvolt}$. We thus conclude that polariton losses are not affected by thermal broadening of the bare excitons.

\begin{figure}[h!]
\centering
    \includegraphics[width=\linewidth]{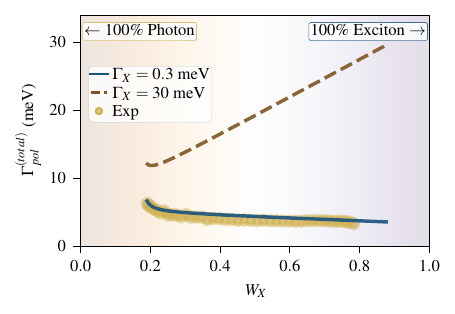}
    \caption{The lines depict the polaritonic linewidths for various excitonic fractions $W_X$, as fitted from Eq.~(1) in the main text. The solid line represents the fit with an excitonic linewidth $\Gamma_X=\SI{0.3}{\milli\electronvolt}$, while the dashed line corresponds to $\Gamma_X=\SI{30}{\milli\electronvolt}$. Scattered data points are derived from experimental PL measurements. 
    }
    \label{fig:loss_suppr}
\end{figure}

\section{Measurement of the group velocity pattern for 63$\%$ excitonic-fraction polaritons}
{With the aim of highlighting the uniform group velocity along the $x$-direction for $y$-polarized polaritons in the 63$\%$ excitonic fraction case, we experimentally measured the group velocity, $v_g$, corresponding to different values of $k_y$. Here, the velocity is extracted from the slope of the dispersion along $k_x$, at various values of $k_y$. Experimentally, the scanning of $k_y$ is achieved by moving the image of the back-focal plane of the microscope objective with respect to the entrance slit of the spectrometer~\cite{Cueff2024}. The results, shown in Fig.~\ref{fig:vel_disp}(a), confirm  that the group velocity does not depend on $k_y$, for both forward and backward propagating polaritons.} 

{Moreover, we can also extract the polariton linewidth, $\Gamma_{pol}$, at different values of $k_y$ from the very same measurements. The results, shown in  Fig.~\ref{fig:vel_disp}(b), confirm that the polariton linewidth does not depend on $k_y$, as predicted by the analytic model.}

\begin{figure}[t]
\centering
    \includegraphics[width=\linewidth]{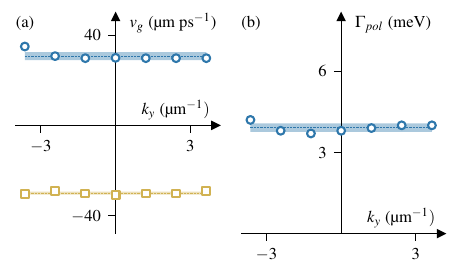}
    \caption{{Experimental results for (a) the group velocity along the $x$ direction, and (b) the linewidth of $y$-polarized polaritons having 63$\%$ excitonic fraction.} The colored areas represent 1 standard deviation around the average values.}
    \label{fig:vel_disp}
\end{figure}

\section{Theoretical model for photonic and polaritonic dispersions}\label{sec:Theory_dispersion}
\subsection{Photonic modes}
{As discussed in detail in Ref.~\cite{art:dang2024}, the two photonic modes that give rise to the two polaritonic branches observed in Figs.~1(c,d) can be described by the following analytic expression:
\begin{equation}\label{eq:photon_disp_full}
\tilde{E}_{ph(\pm)} = E_0+i\frac{\Gamma_0}{2}\pm \sqrt{\left(\hbar vk_x\right)^2+\left(U+i\frac{\Gamma_0}{2}\right)^2 } + \frac{\hbar va}{4\pi}k_y^2 \,  ,
\end{equation}
in which $v$ is the group velocity of the photonic guided modes that are folded and coupled by the periodic metasurface with lattice pitch $a$. The coupling strength between guided modes is quantified by the parameter $U$, while $\Gamma_0$ accounts for the coupling between guided modes and the radiative continuum, due to the periodic dielectric modulation. From this expression, we separate the real part contribution (i.e., the resonances dispersion) from the the imaginary part (i.e., the contribution to the losses):
\begin{equation}
\tilde{E}_{ph(\pm)} =E_{ph(\pm)}+i\frac{\Gamma_{ph(\pm)}}{2} \, .    
\end{equation}
Notice that the $\tilde{E}_{ph+}$ photonic mode in the latter expression corresponds to the mode indicated as $\tilde{E}_{ph}$ in Eq.~(2) of the main text.} 

\subsection{Strong coupling regime}
The coupling between these photonic modes and perovskite excitons gives rise to four polaritonic branches~\cite{art:dang2024}, analytically described as:
\begin{align}\label{eq:Coupled_oscillator_full}
    \tilde{E}_{LP(\pm)}&= \frac{\tilde{E}_{ph(\pm)}+\tilde{E}_X}{2}-\sqrt{V^2+\left[\frac{\tilde{E}_{ph(\pm)}-\tilde{E}_X}{2}\right]^2 } \, ,\\
    \tilde{E}_{UP(\pm)} &= \frac{\tilde{E}_{ph(\pm)}+\tilde{E}_X}{2}+\sqrt{V^2+\left[\frac{\tilde{E}_{ph(\pm)}-\tilde{E}_X}{2}\right]^2 } \, ,
\end{align} 
from which the decomposition into real and imaginary parts can be explicitly written as:
\begin{align}
\tilde{E}_{LP(\pm)} &=E_{LP(\pm)}+i\frac{\Gamma_{LP(\pm)}}{2} \, , \\    
\tilde{E}_{UP(\pm)} &=E_{UP(\pm)}+i\frac{\Gamma_{UP(\pm)}}{2} \, .
\end{align}
Also in this case, notice that the polaritonic mode here indicated as $\tilde{E}_{LP+}$ corresponds to the one denoted as $\tilde{E}_{pol}$ in Eq.~(1) of the main text. The calculated dispersion and linewidth of photonic modes and lower polaritonic modes, obtained from \eqref{eq:photon_disp_full} and \eqref{eq:Coupled_oscillator_full}, are presented in Fig.~\ref{fig:Dispersion_model}. {The model parameters used to obtain the results for these plots read as follows: $a=\SI{300}{\nano\meter}$, $v=\SI[per-mode = symbol]{77.5}{\micro\meter\per\pico\second}$, $U=\SI{10}{\milli\electronvolt}$, $\Gamma_0=\SI{2.2}{\milli\electronvolt} $, $E_X=\SI{2.394}{\electronvolt}$, $E_0=\SI{2.294}{\electronvolt}$, $\Gamma_X= \SI{0.3}{\milli\electronvolt}$ {and $V=\SI{127}{\milli\electronvolt}$}. Finally, to take into account the effect of disorder, an inhomogeneous broadening of $\Gamma_{inh}=3$ meV is added to the polariton linewidths, such that:
\begin{align}
    \Gamma_{LP(\pm)}^\text{(total)}&=\Gamma_{LP(\pm)} + \Gamma_{inh}\, ,\\
    \Gamma_{UP(\pm)}^\text{(total)}&=\Gamma_{UP(\pm)} + \Gamma_{inh}\, .
\end{align}
As shown in the main text, we obtain an almost perfect matching between experimental data and the analytic curves obtained from this model, both for the real and imaginary parts of polaritonic modes, respectively.}
\begin{figure}[t]
\centering
    \includegraphics[width=\linewidth]{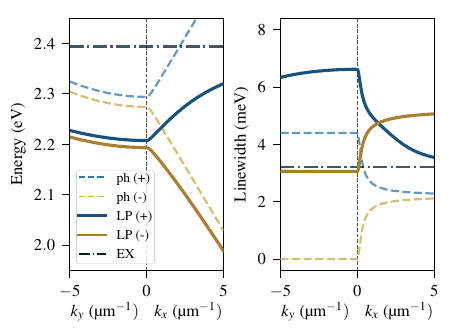}
    \caption{Theoretical calculation of the (a) energy dispersion and (b) linewidth along $k_x$ and $k_y$ of photonic modes $\tilde{E}_{ph(\pm)}$  and lower polaritonic modes $\tilde{E}_{LP(\pm)}$ that are given by \eqref{eq:photon_disp_full} and \eqref{eq:Coupled_oscillator_full}
    , respectively. The parameters are: {$a=\SI{300}{\nano\meter}$, $v=\SI[per-mode = symbol]{77.5}{\micro\meter\per\pico\second}$, $U=\SI{10}{\milli\electronvolt}$, $\Gamma_0=\SI{2.2}{\milli\electronvolt} $, $E_X=\SI{2.394}{\electronvolt}$, $E_0=\SI{2.294}{\electronvolt}$, $V=127$ meV, $\Gamma_X= \SI{0.3}{\milli\electronvolt}$ and $\Gamma_{inh}=3$ meV}}.
    \label{fig:Dispersion_model}
\end{figure}

\subsection{Iso-frequency curves in high momentum limit}
{In the limit of a large momentum, $\hbar v k_x\gg U,\Gamma_0$, the expressions for the photonic modes given in Eq.~\eqref{eq:photon_disp_full} simplify as:
\begin{equation}
\tilde{E}_{ph(\pm)} \approx E_0+i\frac{\Gamma_0}{2}\pm \hbar vk_x+ \frac{\hbar va}{4\pi}k_y^2 \,  .
\end{equation}
Moreover, considering that the losses $\Gamma_0$,$\Gamma_X$ are negligeable with respect to  $\hbar v k_x$ and $V$, the real part dispersion of the lower polariton modes, given  by \eqref{eq:Coupled_oscillator_full}, can be rewritten as:\begin{widetext}
\begin{equation}
E_{LP(\pm)}\approx \frac{E_0+E_X}{2} \pm \frac{\hbar vk_x}{2} + \frac{\hbar va}{8\pi}k_y^2 -\sqrt{V^2+\left[\frac{E_0-E_X}{2} \pm \frac{\hbar vk_x}{2} + \frac{\hbar va}{8\pi}k_y^2\right]^2 } \, .\\
\end{equation}
We notice that the expressions above are for polaritonic modes coming from $E_y$-polarized photonic branches. For polaritonic modes arising from $E_x$-polarized photons, we will have similar expression but switching between $k_x$ and $k_y$. Therefore, the iso-frequency curves at $E_{iso}$ for $LP(+)$ modes are given by:
\begin{align}
\frac{\hbar vk_x}{2} + \frac{\hbar va}{8\pi}k_y^2 -\sqrt{V^2+\left[\frac{E_0-E_X}{2} + \frac{\hbar vk_x}{2} + \frac{\hbar va}{8\pi}k_y^2\right]^2 } + \frac{E_0+E_X}{2} - E_{iso} &=0  \, ,\\
 \frac{\hbar vk_y}{2} + \frac{\hbar va}{8\pi}k_x^2 -\sqrt{V^2+\left[\frac{E_0-E_X}{2} + \frac{\hbar vk_y}{2} + \frac{\hbar va}{8\pi}k_x^2\right]^2 } + \frac{E_0+E_X}{2} - E_{iso} &=0 \, 
\end{align}
for $E_y$ and $E_x$ polarization respectively.
\end{widetext}}

\section{Theoretical model for polariton propagation}\label{sec:Theory_propagation}
{In this section, we present the theoretical model corresponding to the results shown in Fig.~5(d,e,f) of the main text}.

\subsection{Polariton density}
In order to study the propagation of polaritonic excitations within the  metasurface plane, we developed an approximated method to calculate the steady state spatial density distribution of polaritons with complex energy dispersion $\tilde{E}_{pol}=E_{pol} + i\Gamma_{pol}/2$, defined as $n(x,\tilde{E}_{pol})$. {The input parameters of the model are given in the previous Sec.~\ref{sec:Theory_dispersion}}. We discuss only the case of {$y-$polarised polaritons} here, while the case of {$x-$polarised polaritons} can be easily derived from the former. In the case of {$y-$polarised polaritons}, the propagation can be well approximated by a group velocity  pointing in the $x$ direction, $\mathbf{v}_g=v_g\hat{x}$, where $\hat{x}$ is the unit vector of the $x$ coordinate axis. This assumption is justified from the energy cut shown in Fig.~3(a). We start by considering {a generation rate $G(x,E_{pol})$ for polaritons with energy $E_{pol}$ at position $x$, as obtained from a non-resonant} Gaussian excitation spot centered {at $x=0$}, i.e., 
\begin{equation}\label{eq:G}
    G(x,E_{pol})=G_0(E_{pol})e^{-x^2/2\sigma^2 } \, ,    
\end{equation}
in which {the amplitude $G_0(E_{pol})$ represents the spectral distribution of polariton injection}, and $\sigma$ determines the spot size. For a given polariton energy $E_{pol}$, a polaritonic mode has a well defined group velocity $v_g$. {On a first approximation, we can neglect} any nonlinear effect, i.e., we assume that each energy component of the polaritonic dispersion is independent from the others.
In steady state, the total net flux of polaritons is null, and we get the condition:
\begin{equation}\label{eq:flux}
    \frac{\partial n(x,\tilde{E}_{pol})}{\partial x}v_g + \Gamma_{pol}n(x,\tilde{E}_{pol}) = G(x,E_{pol}) \, .
\end{equation}
{Using the generation rate obtained from \eqref{eq:G}, we have the differential equation:}
\begin{equation}\label{eq:eqdiff}
\frac{\partial n(x,\tilde{E}_{pol})}{\partial x}v_g + \Gamma_{pol}n(x,\tilde{E}_{pol}) = G_0(E_{pol})e^{-x^2/2\sigma^2}.
\end{equation}

{At $x=0$, the polariton density is maximized for any eigenmode of energy $E_{pol}$, thus $\frac{\partial n(0,\tilde{E}_{pol})}{\partial x} =0$, which { gives the boundary condition:}
\begin{equation} \label{eq:boundary}
 n(0,\tilde{E}_{pol}) = \frac{G_0(E_{pol})}{\Gamma_{pol}} \, .
\end{equation}}


{Moreover, for each polariton eigenmode with complex energy $\tilde{E}_{pol}$, using \eqref{eq:boundary} as initial value at $x=0$, \eqref{eq:eqdiff} } can be easily integrated numerically, which gives the solution $n(x,\tilde{E}_{\text{pol}})$ {of the polariton density}. 

{We finally notice that the same excitation spot is also responsible for the injection of non-propagating uncoupled excitons with density: 
\begin{equation}
n_X(x)=n_0e^{-x^2/2\sigma^2} \, .    
\end{equation}}

\subsection{The PL signal}
{The photoluminesce (PL) spectrum measured at energy $E$ and due to a polariton eigenmode with complex energy $\tilde{E}_{pol} = E_{pol} + i\Gamma_{pol}/2$ can be modeled by the following Lorentzian profile:
 \begin{equation}
    S_{pol}(E,\tilde{E}_{pol})=\frac{(\Gamma_{pol} + \Gamma_{inh}) }{\pi \left[ (E-E_{pol})^2+(\Gamma_{pol} + \Gamma_{inh})^2/4 \right]} \, .
\end{equation}
The PL intensity of the polariton mode at energy $E_{pol}$, when spatially selected at position $x\in [x_1,x_2]$, is given by:
\small\begin{equation}\label{eq:intensity}
    I_{pol}(E,\tilde{E}_{pol})\propto  \underbrace{S_{pol}(E,\tilde{E}_{pol})}_\text{spectrum}\underbrace{(\Gamma_{pol}-W_X\Gamma_X)}_\text{radiative losses}\!\underbrace{\int_{x_1}^{x_2} n(x,\tilde{E}_{pol})dx}_\text{polariton population}.
\end{equation}
\normalsize

Similarly, the PL spectrum and the PL intensity of uncoupled excitons, when spatially selected in position $x\in [x_1,x_2]$, are respectively given by:{
\begin{align}
    S_X(E)&=\frac{\Gamma_X^\text{(uncoupled)} }{\pi \left[ (E-E_X)^2+\left(\Gamma_X^\text{(uncoupled)}\right)^2/4 \right]}\\
    I_X(E)&\propto S_X(E)\int_{x_1}^{x_2} n_X(x)dx \, .
\end{align}
Here, the linewidth $\Gamma_X^\text{(uncoupled)}$ of uncoupled excitons is the room temperature one, i.e., including thermal broadening effects, and thus $\Gamma_X^\text{(uncoupled)}= 30$\,meV.}}

{Taking into account the two lower polaritonic branches, $LP(\pm)$, and the uncoupled excitons, the total PL intensity at energy $E$ and wavevector $k_x$, when spatially selected at position $x\in [x_1,x_2]$, is given by:
\begin{equation} \label{eq:I_total}
    I_{total}(E,k_x) = \sum_{pol=LP(\pm)} I_{pol}\left(E,\tilde{E}_{pol}(k_x)\right) + I_X(E) \, ,
\end{equation}
in which $E_{LP(\pm)} (k_x)$ represent the dispersions of the lower polariton modes, respectively, as given in \eqref{eq:Coupled_oscillator_full}. }
{Finally, a selection in real space (i.e., spatial selection) induces a broadening in momentum space, due to the reciprocal relation $\Delta_x \cdot \Delta_k\sim 1$. This broadening is taken into account by a convolution of $ I_{total}(E,k_x)$ with a Gaussian function whose standard deviation is $\Delta_k$.}

\subsection{Numerical parameters}
{The dependence of the polariton energy on the wave-vector along the propagation direction, $E_{pol} (k_x)$, and its corresponding linewidth, $\Gamma_{pol}(k_x)$, is calculated as detailed in  Sec.~\ref{sec:Theory_dispersion},  for both polaritonic eigenmodes. The only supplemental parameters to calculate the spatial distribution of polariton density [see \eqref{eq:flux}] and the PL in momentum space [see \eqref{eq:I_total}] are the spectral distribution $G_0(E_{pol})$ for the polariton injection, and the injection density $n_0$ for uncoupled excitons. {For the former, we apply a simple phenomenological law, i.e., assuming $G_{0}(E_{pol})\propto\exp\left(-|E_{pol}-E_{c}|/W\right)$, which exhibits a polariton concentration at energy $E_{c}$. Numerically, we assume $W=\SI{30}{\milli\electronvolt}$,  $E_{c}=\SI{2.209}{\electronvolt}$ for LP(+) mode and  $E_{bn}=\SI{2.17}{\electronvolt}$ for LP(-) mode. For the amplitude of injection, we assume that the LP(+) mode is pumped 80 times more efficiently than the LP(-) one}. Using these parameters, the theoretical calculation for the decay of 63$\%$ excitonic-fraction polaritons in real-space, as well as the polariton density in momentum-space after spatial selection, are presented in Fig.~4 in the main text. We highlight the remarkably good agreement between these calculation and the experimental data presented in Figs.~4 and 5 of the main text. }

\begin{figure}[t] 
\centering
    \includegraphics[width=0.85\linewidth]{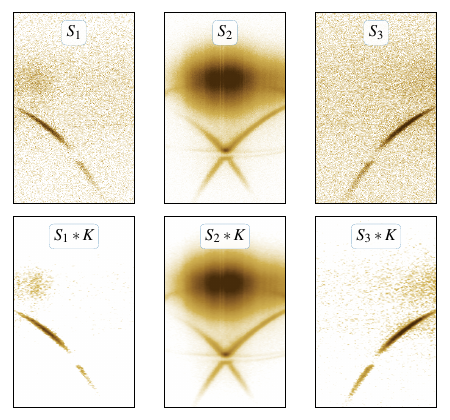}
    \caption{{$S_1$, $S_2$, and $S_3$ are three PL signals coming from three different positions on the sample. These signals are then convoluted using the kernel $K$ defined in~\eqref{eq:gavss}, and reported in the plot as $S_1\ast K$, $S_2\ast K$ and $S_3\ast K$. These convoluted signals are the ones also reported in Fig.~5(a-c) of the main text.}}
    \label{fig:supp_conv}
\end{figure}
\begin{figure*}[!ht]
\centering
    \includegraphics[width=0.96\linewidth]{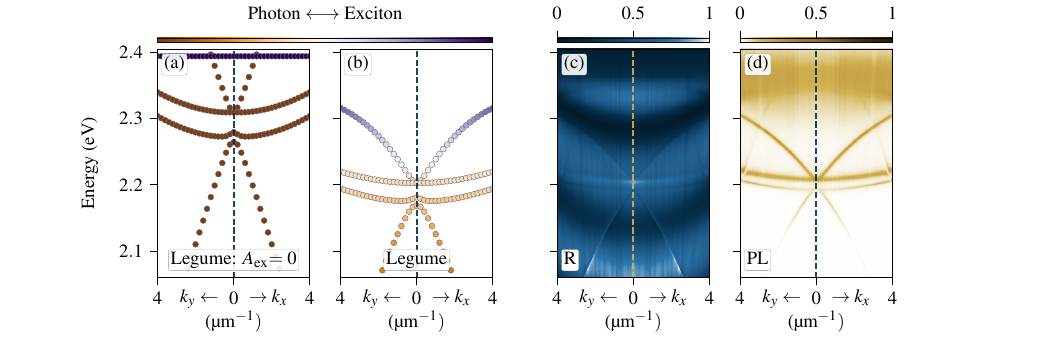}
    \caption{{(a-b) Polariton dispersion calculated with \texttt{legume}, with quenched and normal oscillator strength, respectively. (c-d) Experimentally measured unpolarized angle-resolved reflectivity and PL spectra.}  }.
    \label{fig:weak_strong}
\end{figure*}

\section{Denoising of PL signal}
The signal-to-noise ratio of propagated polaritons in Fig.~5(a-c) is particularly low. This is due to the signal collection being performed at $|x|=\SI{100}{\micro\meter}$ from the excitation spot, while the propagation length fitted from Fig.~4 is estimated to be of the order of $l\approx \SI{20}{\micro\meter}$. To ``denoise'' the experimental data of Fig.~5, we performed a convolution $S\ast K$ of the signal $S$ using a $5\times5$ Gaussian kernel $K$ defined as:
\begin{equation}\label{eq:gavss}
    K=\frac{1}{2^8} \begin{bmatrix}
1 & 4 & 6 & 4 & 1 \\
4 & 16 & 24 & 16 & 4 \\
6 & 24 & 36 & 24 & 6 \\
4 & 16 & 24 & 16 & 4\\
1 & 4 & 6 & 4 & 1
\end{bmatrix} \,.
\end{equation}
The convolution kernel in~\eqref{eq:gavss} is normalized such that the sum of all its elements is equal to 1. Since the Fourier transform of a Gaussian is still a Gaussian, the convolution between the actual signal and $K$ quenches the high-frequency components. The low-pass filtering effect of such a convolution effectively increases the signal-to-noise ratio. As an example, this is explicitly demonstrated in {Fig.~\ref{fig:supp_conv}, in which we show three the signals obtained from PL measurements before (upper panels) and after (lower panels) the convolution.} {The same convolution technique has been used to denoise the experimental data shown in Fig.~3 in the main text.}

\section{METHODS}
\subsection{Sample fabrication}
{ The fabrication process initiates by {depositing} a thin layer of PEPI onto substrates composed of \SI{2}{\micro\meter} thick SiO$_2$ on silicon (Si). This is achieved by spin-coating a 20\% wet solution of PEPI in Dimethylformamide (DMF) at 5000 rpm for 30 seconds. Following this step, the PEPI film undergoes annealing at 95°C for 90 seconds to induce crystallization before proceeding to the imprinting step. {Subsequently}, a Si mold with the desired structure is pressed onto the PEPI film using a thermal press. This imprinting process occurs at 100°C under a pressure of 100 bar for approximately 10 minutes. As a result, the structure from the mold is directly transferred onto the PEPI film. {Details of the fabrication method were already reported in Ref.~~\cite{art:dang2024}.} }
 
\subsection{Experimental setup}
{The sample is illuminated via a microscope objective (20x magnification, NA=0.42) using a white-light beam (halogen lamp) for ARR experiments or a non-resonant laser ($\SI{405}{\nano\meter} $, \SI{80}{\mega\hertz}, \SI{50}{\pico\second}) for ARPL experiments . The signal that is scattered or emitted from the sample is then collected through the same objective, and analyzed  by imaging the back-focal plane of the microscope objective for measurements in momentum-space, or the sample plane for measurements in real-space. For dispersion measurements, the signal is first projected onto the entrance of a spectrometer, whose output is coupled to the sensor of a CCD camera. For the propagation experiment in real-space, the signal is spectrally filtered by a band pass filter at {543$\pm 5$~nm (Thorlabs FL543.5-10)} , and then projected onto the sensor of a SCMOS camera. For the propagation experiment in momentum space, the signal is spatially filtered by a slit that is positioned in correspondence of the plane of the intermediate image~\cite{Cueff2024}}.

{ 
\section{Legume simulation of active metasurface}\label{sec:legume_params}
In Fig.~1(c-e) we compare the experimental spectra with the polariton bands calculated using the free software \texttt{legume}~\cite{art:Zanotti2024,repo:legume}. The simulated metasurface is shown in Fig.~1(a) of the manuscript. The main structural parameters are: lattice constant $a=\SI{295}{\nano\meter}$, the total thickness of the PEPI layer $t_\text{PEPI}=\SI{93}{\nano\meter}$, etching depth $t_\text{et}=\SI{35}{\nano\meter}$, the thickness of the PMMA layer $t_\text{PEPI}=\SI{182}{\nano\meter}$, and the hole radius $r=\SI{125}{\nano\meter}$. The refractive indices of the materials used in the simulations are assumed as: $n_{\ch{SiO2}}=1.46$, $n_\text{PMMA}=1.46$, $n_{\text{PEPI}}=2.4$. The excitons are modeled using 10 two-dimensional effective active layers. These active layers are placed in $N_z=5$ equally spaced positions in the PEPI region. Each position is assumed to have both an $x-$ and $y-$polarized active layer (i.e., exciton species). The exciton oscillator strength per unit surface of each active layer is given by the expression:
\begin{equation}
    \frac{f}{S}=\frac{\varepsilon_0 m_0 A_\text{ex} }{\hbar^2} L
\end{equation}
where $\varepsilon_0$ is the vacuum dielectric permittivity, $m_0$ is the electron mass,  $L=t_\text{PEPI}/N_z$ is the thickness of the effective active layer, and $A_\text{ex}=\SI{0.85}{\square\electronvolt}$ (see, e.g., Ref.~\cite{Dang2020}). 
To further investigate the role of PEPI as the active material, we performed two simulations. In the first one, we artificially set the oscillator strength to $A_\text{ex}=\SI{0}{\square\electronvolt}$, thus completely decoupling excitons from photons, as shown in Fig.~\ref{fig:weak_strong}(a). In Fig.~\ref{fig:weak_strong}(b), the oscillator strength is restored to $A_\text{ex}=\SI{0.85}{\electronvolt}$, leading to the same exciton-polariton dispersion visible in the experimental spectra of Fig.~\ref{fig:weak_strong}(c-d). These results confirm that our sample is indeed in the strong coupling regime.
}

{
\begin{figure}[t]
\centering
    \includegraphics[width=\linewidth]{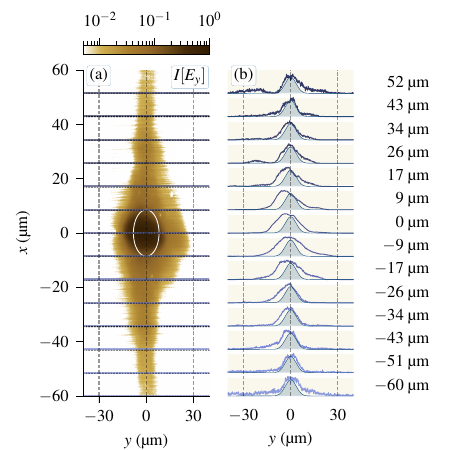}
    \caption{(a) Spatially resolved PL intensity of $E_y$-polarized polaritons at $E=\SI{2.283}{\electronvolt}$. (b) Cuts of the intensity profile in (a) at different position $x$ indicated on the right. The blue shaded area corresponds to the pump-spot Gaussian profile. Each cut is individually normalized.}.
    \label{fig:Cuts}
\end{figure}
\section{Polariton propagation vs. exciton diffusion}
In this section, we further investigate the polariton propagation both in real and momentum space. We start by considering the diffusion of excitons in 2D hybrid perovskites in real space, which has been already quantified to be in the few hundred nanometers range~\cite{Ziegler2020, Seitz2020, Xiao2020}. Since uncoupled excitons are injected within the area of the pump spot on the sample, exciton diffusion is unequivocally ruled out for the long-range propagation far away from the pump spot.
}

{
On the other hand, to highlight exciton diffusion in our sample, one may monitor the broadening of the spatial profile of the PL signal along the direction perpendicular to that of polariton propagation. Figure~\ref{fig:Cuts}(b) shows the PL profile along the $y$ direction, taken at various $x$ locations of polariton propagation as shown in Fig.~4 of the manuscript (corresponding to Fig.~\ref{fig:Cuts}(a)). In this measurement, a spectral filter at \SI{2.283}{\electronvolt} is applied, corresponding to polaritons with a 63$\%$ excitonic fraction and a small part of uncoupled excitons in the emission tail. These results show that the $y-$profile emission is identical to the pump spot profile when taken far from the pump spot region ($x\sim\SI{50}{\micro\meter}$), which is justified by polaritonic unidirectional propagation. On the contrary, within the tail of the pump spot, the $y-$profile of the emission is broadened as a result of exciton diffusion.  }

\begin{figure}[t]
\centering
    \includegraphics[width=\linewidth]{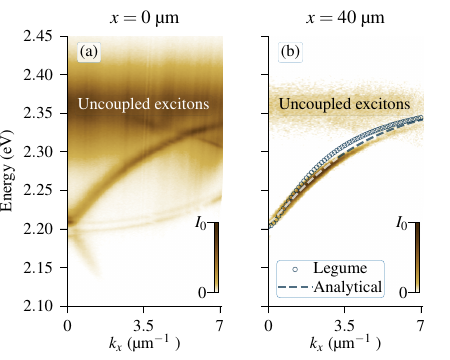}
    \caption{{(a) PL intensity filtered around the pump spot at $x=\SI{0}{\micro\meter}$. The signal is strongly dominated from the emission of uncoupled excitons below the the bare exciton energy $E_X=\SI{2.394}{\electronvolt}$. (b) PL intensity filtered at $\SI{40}{\micro\meter}$ from the pump spot where the signal from uncoupled exciton is strongly reduced. The signal from the LP mode is compared with the dispersion calculated from \texttt{legume} and the analytical model.}}.
    \label{fig:real_filters}
\end{figure}
{
In addition, exciton diffusion can also be observed in spectrally resolved PL measurements in momentum space. Figure~\ref{fig:real_filters} shows the angle-resolved PL filtered at $x=\SI{0}{\micro\meter}$ and at $x=\SI{40}{\micro\meter}$, respectively. The spatial width of the applied filter is \SI{17}{\micro\meter}.  These results clearly show that the emission within the pump spot is dominated by uncoupled excitons. On the contrary, the emission outside the pump spot mostly comes from the LP band, which possesses positive group velocity. However, a very weak emission of uncoupled excitons is still visible, and it is attributed to the emission of propagated excitons via exciton diffusion.
}
\section{$E_x$-polarized polaritons propagation }
Throughout this work, we focused our attention on the $E_y$-polarized polaritons propagating along the $x$ direction. These propagating polaritons are easily identified in Fig.~\ref{fig:real_filters}(b). We have already seen the corresponding PL measurement filtered at $E=\SI{2.283}{\electronvolt}$ for $E_y$-polarized polaritons, which is reported in Fig.~\ref{fig:Cuts}(a), clearly showing directional propagation along $x$. Due to the $C_4$ symmetry of the square lattice, the same propagation is expected along the $y$ direction, for $E_x$-polarized polaritons. Here we report this measurement for completeness, which confirms the expected result, as shown in Fig.~\ref{fig:y_propagation}. The measurement has been performed several months later than the one shown in Fig.~\ref{fig:Cuts}(a), which justifies the lower quality of the image.
\begin{figure}[t]
\centering
    \includegraphics[width=\linewidth]{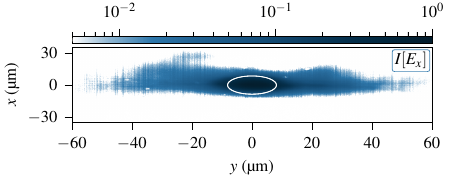}
    \caption{Spatially resolved PL intensity of $E_x$-polarized polaritons propagating along $y$, filtered at $E=\SI{2.283}{\electronvolt}$.}.
    \label{fig:y_propagation}
\end{figure}

\newpage


\end{document}